\documentclass[11pt]{book}
\usepackage{amsmath,epsfig,graphicx,amssymb}
\begin{document}
    \setlength{\textheight}{8in} 
    \setlength{\textwidth}{5.5in}
    \setlength{\oddsidemargin}{0.5in}
    \setlength{\evensidemargin}{0.5in}

\pagestyle{myheadings} 
    \markboth{E. Silva, N. Pompeo, S. Sarti, C. Amabile}{Vortex state
    microwave response in superconducting cuprates and MgB$_{2}$}
    \renewcommand{\chaptername}{\vskip-20mm\protect
    \normalsize\protect\huge\bf Chapter}

\setcounter{chapter}{0} \chapter{Vortex state microwave response in
superconducting cuprates and MgB$_{2}$} \setcounter{page}{1}

\begin{center}

{ \large {\bf E. Silva$^\dag$, N. Pompeo$^\ddag$, S. Sarti$^\S$, C.
Amabile$^{\P}$} } \vspace{3mm}

{ \large {\it $^\dag$ Dipartimento di Fisica ``E. Amaldi'' and Unit\`a
CNISM,\\
Universit\`{a} Roma Tre, Via della Vasca Navale 84, 00146 Roma,
Italy\\
\vspace{3mm}
{e-mail: silva@fis.uniroma3.it}}
}
\vspace{10mm}

{
\large {\it $^\ddag$ Dipartimento di Fisica ``E.Amaldi'' and Unit\`a CNISM,
Universit\`{a} Roma Tre, Via della Vasca Navale 84, 00146 Roma,
Italy\\
}
}
\vspace{10mm}

{
\large {\it $^\S$ Dipartimento di Fisica and Unit\`a CNISM,\\
Universit\`{a} ``La Sapienza'', P.le Aldo Moro 2, 00185 Roma,
Italy\\ \vspace{3mm} {e-mail: stefano.sarti@roma1.infn.it}}
}
\vspace{10mm}

{
\large {\it $^\P$ Dipartimento di Fisica,\\
Universit\`{a} ``La Sapienza'', P.le Aldo Moro 2, 00185 Roma,
Italy\\
}
}
\vspace{10mm}

\begin{minipage}{5in}
\centerline{{\sc Abstract}}
\medskip

We investigate the physics of the microwave response in
YBa$_{2}$Cu$_{3}$O$_{7-\delta}$, SmBa$_{2}$Cu$_{3}$O$_{7-\delta}$ and
MgB$_{2}$ in the vortex state.  We first recall the theoretical basics
of vortex-state microwave response in the London limit.  We then
present a wide set of measurements of the field, temperature, and
frequency dependences of the vortex state microwave complex
resistivity in superconducting thin films, measured by a resonant
cavity and by swept-frequency Corbino disk.  The combination of these
techniques allows for a comprehensive description of the microwave
response in the vortex state in these innovative superconductors.  In
all materials investigated we show that flux motion alone cannot take
into account all the observed experimental features, neither in the
frequency nor in the field dependence.  The discrepancy can be
resolved by considering the (usually neglected) contribution of
quasiparticles to the response in the vortex state.  The peculiar,
albeit different, physics of the superconducting materials here
considered, namely two-band superconductivity in MgB$_{2}$ and
superconducting gap with lines of nodes in cuprates, give rise to a
substantially increased contribution of quasiparticles to the
field-dependent microwave response.  With careful combined analysis of
the data it is possible to extract or infer many interesting
quantities related to the vortex state, such as the
temperature-dependent characteristic vortex frequency and vortex
viscosity, the field dependence of the quasiparticle density, the
temperature dependence of the $\sigma$-band superfluid density in
MgB$_{2}$

\end{minipage}
\end{center}

\noindent {\bf Keywords:} superconductivity, surface impedance,
microwaves, vortex motion, cuprates, YBa$_{2}$Cu$_{3}$O$_{7-\delta}$,
SmBa$_{2}$Cu$_{3}$O$_{7-\delta}$, MgB$_{2}$, vortex viscosity.

\section{INTRODUCTION}
\label{intro}
One of the most versatile experimental methods for the investigation
of the physics of superconductors is the measurement of the complex
response to an alternating electromagnetic (e.m.) field in the
radiofrequency (\textit{rf}) and microwave ranges.  The resulting data
have been of great help in the understanding of the physics of
conventional superconductors.  Even confining the treatment to the
linear response, in conventional superconductors microwave or
\textit{rf} measurements allowed, e.g., for the determination of the
existence \cite{BiondiRMP58} and temperature dependence
\cite{BiondiPR59} of the superconducting gap, for the settling of the
kind of dynamical fluctuations
\cite{DaielloPRL69,LehoczkyPRL69,LehoczkyPRB71}, for the determination
of the penetration depth \cite{PippardPRS47,PippardPRS50,WaldramAP64},
for the determination of the thermodynamical critical field and of the
third critical field for surface superconductivity
\cite{RosenblumPL64}, and for the determination of the upper critical
field \cite{RosenblumPRL64}.

In type-II superconductors the investigation was extended to the mixed
state: when the magnetic field $H_{c1}<H<H_{c2}$, with $H_{c1}$ and
$H_{c2}$ the temperature dependent lower and upper critical field,
respectively, the magnetic flux penetrates the superconductor as
quantized flux lines, each carrying one flux quantum
$\Phi_{0}=$2.07$\times$10$^{-15}$ T$\cdot$m$^{2}$.  Such flux lines,
in presence of an electric current density $\mathbf{J}$, are subjected
to a Lorentz force (per unit length) $\mathbf{F_{L}} = \mathbf{J}
\times \mathbf{\Phi_{0}}$ (where $\mathbf{\Phi_{0}}\parallel
\mathbf{B}$).  Moving vortices dissipate energy, so that an ideal
type-II superconductor always has a finite resistivity in the vortex
state.  Dissipation can be viewed either as due to the continuous
conversion of Cooper pairs into quasiparticles at the (moving) vortex
boundary \cite{TinkhamINTROSUP}, or to Joule heating inside of the
vortex core \cite{Abrikosov,BardeenPR65}.  In any case, the
dissipation is expected to depend mainly on the fundamental
properties of the superconductors, such as the quasiparticle density
of states and relaxation time in the vortex core, and thus it is
expected to be very similar in different samples of the same
material.

A dissipationless regime can be achieved in dc by pinning vortices to
defects \cite{Degennes,Kim69,CampbellAP72}.  When the force acting on
vortices is alternating, pinning determines an additional imaginary
(out-of-phase) response.  As opposed to energy dissipation, the
particular pinning mechanism and its efficacy are expected to be
strongly sample dependent.

To take into account the energy dissipation and the effects of pinning
a simple way is to build up an equation of motion for a single vortex.
In a simple model, independently pinned vortices are considered and
the equation of motion for the displacement $\mathbf{u}$ from the
equilibrium position is written in the elastic
approximation\footnote{The vortex mass is usually neglected
\cite{SuhlPRL65,MatsudaPRB94}, even if this is still a debated topic 
\cite{KopninPRL98,SoninPRB01,HanPRB05}.}:
$\eta\mathbf{\dot{u}}+\kappa_{p}\mathbf{u}=\mathbf{F_{L}}$, where
$\eta$ is the so-called {\it vortex viscosity} (per vortex unit
length) and takes into account the energy dissipation, and
$\kappa_{p}$ is the {\it pinning constant} (per vortex unit length)
and takes into account the pinning of vortices.  This simple equation
of motion gives rise to a complex vortex motion resistivity due to an
alternating current density $\mathbf{J}e^{\mathrm{i}\omega t}$, with
$\nu=\omega/2\pi$ the measuring frequency, that can be written as
\cite{GittlemanPRL66}:
\begin{equation}
\rho_{vGR}=\frac{\Phi_{0}B}{\eta}\frac{1}{1-\mathrm{i}\frac{\kappa_{p}}{\eta\omega}}=
      \rho_{ff}\frac{1}{1-\mathrm{i}\frac{\omega_{p}}{\omega}}
      \label{semplice}
\end{equation}
where the last equality defines the so-called {\it flux-flow
resistivity}, $\rho_{ff}$, and the {\it depinning frequency}
(sometimes called ``pinning frequency'') $\nu_{p}=\omega_{p}/2\pi$.
Referring to the physical meaning of the vortex parameters, $\eta$ is
expected to be a very similar, if not universal, function of the
external parameters (e.g., the temperature) in different samples of
the same material: it is a quantity related to the fundamentals
of the physics of type-II superconductors.  In particular, it was
demonstrated by Bardeen and Stephen (BS) \cite{BardeenPR65} that, in
dirty superconductors, $\eta=\Phi_{0}B_{c2}/\rho_{n}$, where
$\rho_{n}$ is the normal state dc resistivity and $B_{c2}$ is the
upper critical field.

In deep contrast, one expects largely
different magnitudes, temperature and magnetic field dependence for
$\kappa_{p}$ (or $\omega_{p}$) in different samples even of the same
material, being the pinning constant related to
extrinsic properties.  Its study can, on
one side, shed light on the characteristic features of the vortex
matter, and on the other side can be of essential importance for the
applications of superconductors \cite{Degennes,Kim69,CampbellAP72}.

\begin{figure}[t]
\begin{center}
\begin{minipage}[h]{80mm}
\epsfig{file=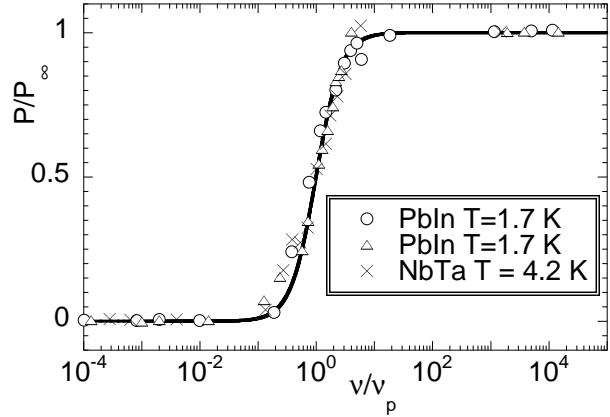, width=80mm}
\end{minipage}
\caption{Normalized dissipated power in type-II superconductors at
$H=\frac{1}{2}H_{c2}$ as a function of the measuring frequency
normalized to $\nu_{p}$ (replotted from \cite{GittlemanPRL66}) and fit
by Eq.(\ref{semplice}).  $\nu_{p}=$ 3.9, 5.1, 15 MHz
\cite{GittlemanPRL66}.}
\label{gittrosdata}
\end{center}
\end{figure}
\begin{figure}[t]
\begin{center}
\begin{minipage}[h]{80mm}
\epsfig{file=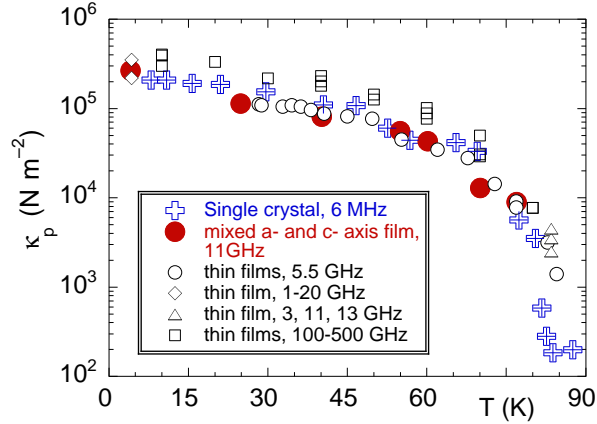, width=80mm}
\end{minipage}
\caption{Pinning constant $\kappa_{p}$ for many different YBCO
samples: \cite{WuPRL90}, crosses; \cite{Pambianchi93}, red full dots;
\cite{GolosovskyPRB94}, open circles; \cite{RevenazPRB94}, diamonds;
\cite{ParksPRL95}, open squares; \cite{WuPRL95}, triangles.  Data
replotted from original papers.  A remarkable collapse of the data on
a single curve is evident.}
\label{goloskappa}
\end{center}
\end{figure}
\begin{figure}[t]
\begin{center}
\begin{minipage}[h]{80mm}
\epsfig{file=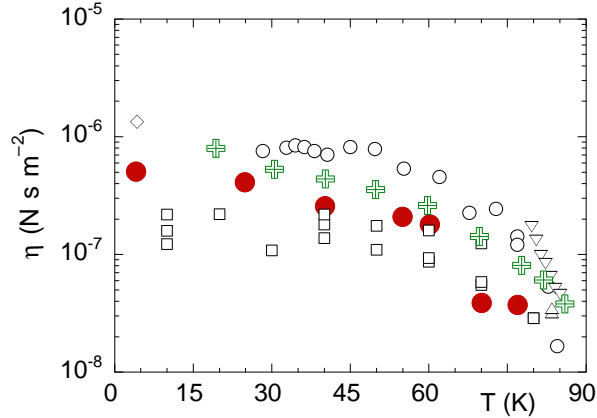, width=80mm}
\end{minipage}
\caption{Vortex viscosity $\eta$ for many different YBCO samples:
\cite{OwliaeiPRL92}, down triangles; \cite{Pambianchi93}, red full
dots; \cite{GolosovskyPRB94}, open circles; \cite{RevenazPRB94},
diamonds;\cite{MorganPhC94}, crosses; \cite{ParksPRL95}, open squares;
\cite{WuPRL95}, triangles.  Data replotted from original papers.  A
remarkable spread of the data is evident.}
\label{goloseta}
\end{center}
\end{figure}

The applicability of Eq.(\ref{semplice}) was first brought to the
attention of the scientific community with the seminal paper by
Gittleman and Rosenblum \cite{GittlemanPRL66}: as shown in
Fig.\ref{gittrosdata}, the data for the dissipated power
($\propto\mathrm{Re}[\rho_{vGR}]$) in thin superconducting films
followed very closely Eq.(\ref{semplice}) over several orders of
magnitude for the measuring frequency.  The success of this extremely
simple model determined its assumption for the interpretation of the
data taken in high-$T_{c}$ superconductors (HTCS).  Many studies were
carried out in HTCS, aimed at the determination of the vortex
viscosity and pinning constant through measurements of the
radiofrequency, microwave or millimeter-wave response
\cite{WuPRL90,YehPRB91,GolosovskyPRB92,HuangPhC92,OwliaeiPRL92,Pambianchi93,SilvaPhC93b,
GolosovskyPRB94,RevenazPRB94,MorganPhC94,ParksPRL95,WuPRL95,
GolosovskySUST96,BelkPRB96,BelkPRB97,LutkePRL97,OngPRB97,HanaguriPRL99,
RogaiIJMPB00, SilvaSUST00,TsuchiyaPRB01,SilvaPhC04}.  In particular,
in the most studied compound YBa$_{2}$Cu$_{3}$O$_{7-\delta}$ (YBCO),
it was found that the depinning frequency was located in the GHz
range, thus making microwave techniques the ideal candidate method for
the experimental investigation of the vortex motion.  However, even
after many years from the discovery of YBCO, the values given for
vortex parameters as determined by microwave measurements (see
\cite{GolosovskySUST96} for a clear overview) presented a very
puzzling framework.  As reported in Figs.\ref{goloskappa} and
\ref{goloseta}, with data replotted from original papers
\cite{WuPRL90,OwliaeiPRL92,Pambianchi93,
GolosovskyPRB94,RevenazPRB94,MorganPhC94,ParksPRL95,WuPRL95} cited in
\cite{GolosovskySUST96}, it was shown that while data for $\kappa_{p}$
taken on different samples (c-axis and mixed a- and c- axis films from
various sources, single crystals) lie all together on the same curve,
the vortex viscosity presented a wide scattering of the data.  As
discussed above, due to the meaning of the vortex parameters this
finding is not easily explained: one would rather expect, for a given
material, a universal behavior for $\eta$ and a sample-dependent
$k_p$.

Similar oddities were systematically evident also in different
frequency ranges in various HTCS. Sub-THz measurements of the
vortex-state microwave resistivity of thin YBCO films
\cite{ParksPRL95} showed that vortex parameters obtained from
conventional, vortex-motion-driven response, were in strong contrast
with calculations of the same parameters from microscopic theories.
Again, the pinning constant was found nearly sample-independent.  The
apparent vortex viscosity would differ from the microscopic
calculation by more than an order of magnitude.  An alternative
analysis of the data \cite{ParksJPCS95} suggested that it was not
possible to ignore, in the interpretation of the data, the effect of
field-induced pair breaking.  Accordingly to that picture, data of the
imaginary conductivity in the same frequency range in
Bi$_{2}$Sr$_{2}$CaCu$_{2}$O$_{8+x}$ (BSCCO) films in the vortex state
\cite{MallozziPRL98} could be accounted for even without resorting to
any vortex motion model, being the field-induced superfluid density
suppression sufficient for a quantitative description.

Summarizing, it was clear that Eq.(\ref{semplice}) did not provide a
comprehensive explanation of the vortex state microwave response in
HTCS. Even if much more complicated (and probably more realistic)
vortex models could be invoked \cite{BlatterRMP94,BrandtRPP95},
additional mechanisms had to be investigated.

Granularity has been sometimes indicated as a possible dominant source
for the losses in the microwave response in superconducting films.
Manifestations of granularity include weak-links dephasing
\cite{PakulisPRB88,MarconPRB89,GiuraPRB89,GiuraPRB90,WosikJAP91,WosikPRB93,Fastampa96},
Josephson fluxon (JF) dynamics \cite{HalbritterJS95} and, as recently
studied, Abrikosov-Josephson fluxon (AJF) dynamics
\cite{GurevichPRB92,GurevichPRB02}.  Weak-links dephasing is
charaterized by a very sharp increase of the dissipation at dc fields
of order of, or less than, 20 mT, accompanied by a strong (and
sometimes exceptionally strong) hysteresis with increasing or
decreasing field
\cite{PakulisPRB88,MarconPRB89,GiuraPRB89,GiuraPRB90}.  However, those
effects are relevant in large-angle grain boundaries (or very weak
links), such as those found in pellets and granular samples, and are
not observed in good thin films.  Josephson fluxon dynamics has been
studied essentially in relation to nonlinear effects
\cite{XinPRB02,LeePRB05}, due to the short JF nucleation time.  One of
the characteristic features of AJF should be a rather pronounced
sample dependence (defect-driven) of the microwave response, as in
fact reported in Tl-2212 films \cite{GaganidzeJAP03}.  This fact
contrasts with the pseudo-universal behavior of the estimated
$\kappa_{p}$, even if contributions from AJF cannot in general be
excluded {\it a priori}.

Another intrinsic, unavoidable source for the measured microwave
resistivity is obviously the conductivity due to charge carriers, that
we write in terms of the so-called ``two-fluid model''
\cite{TinkhamINTROSUP}: charge carriers are thought as given by a
superconducting fraction $x_{s}$ and a ``normal'' fraction (more
appropriately, due to quasiparticle excitations) $x_{n}$.  When the
measuring frequency is much smaller than the quasiparticle relaxation
rate this leads to the well-known dependence of the charge
carriers conductivity at nonzero frequency:
\begin{equation}
      \sigma=\sigma_{1}-\mathrm{i}\sigma_{2}=
      \frac{ne^{2}}{m\omega}\left(\omega\tau_{qp}x_{n}-\mathrm{i}x_{s}\right)=
\frac{1}{\mu_{0}\omega}\left(\frac{2}{\delta^{2}_{nf}}-\mathrm{i}\frac{1}{\lambda^{2}}\right)
      \label{twofluid}
\end{equation}
where $n$ is the charge carrier density, $m$ is the carrier effective
mass, $\tau_{qp}$ is the quasiparticle scattering time, and the last
equality defines the temperature, field and frequency dependent normal
fluid skin depth $\delta_{nf}$ and the temperature and field dependent
London penetration depth $\lambda$. How this charge carriers 
conductivity combines with the vortex motion resistivity is discussed 
in Section \ref{theory}.

It should be mentioned that measurements of $\sigma_{1}$ and
$\sigma_{2}$ in zero applied field have been demonstrated to be of
crucial importance in the determination of the peculiarities of HTCS.
As a partial list of examples, microwave measurements of the
low-temperature variation of the superfluid fractional density,
expressed as
$x_{s}(T)=\left[\frac{\lambda(0)}{\lambda(T)}\right]^{2}$, have shown
a clear nonactivated behavior in YBCO \cite{HardyPRL93,MaoPRB95},
thus giving strong evidence for an anisotropic superconducting gap,
with lines of nodes.  In particular, a linear decrease with
temperature was found in clean crystals \cite{HardyPRL93,MaoPRB95},
demonstrating the existence of zero-energy excitations at low
temperatures.  Moreover, it was shown by microwave spectroscopy of the
real part of the response in YBCO \cite{BonnPRB93} that the
quasiparticle relaxation rate $1/\tau_{qp}$ decreased, with respect to
its value above $T_{c}$, by up to two orders of magnitudes.  Short
relaxation rates below $T_{c}$ were confirmed in YBCO films by
millimeter-wave interferometry \cite{NagashimaFujita94}, in YBCO
crystals by surface resistance measurements \cite{JacobsJPCS95}, and
in YBCO films by far-infrared measurements \cite{GaoPRB96}.  Crystals
consistently presented $1/\tau_{qp}$ at low temperatures up to two
order of magnitudes smaller than in the normal state
\cite{BonnPRB93,JacobsJPCS95}.  Similar drops were found in some films
\cite{NagashimaFujita94}, while other films presented a difference in
$1/\tau_{qp}$ of only a factor of two \cite{GaoPRB96}.

All the above mentioned mechanisms present novel features in the
recently discovered \cite{NagamatsuNAT} metallic superconductor
MgB$_{2}$.  The well-established, albeit sensitive to interband
scattering, two-gap nature of this compound
\cite{BouquetPRL01,IavaronePRL02,GonnelliPRL02} is expected to
strongly affect the intrinsic properties, and specifically to the
present study, the vortex viscosity and the superfluid density.
Microwave measurements \cite{KimPRB02,JinPRB02,GhigoPRB05} in zero dc
magnetic field showed peculiar temperature dependences of the surface
impedance, due to the existence of the double gap.  Measurements of
the magnetic-field-dependent microwave surface impedance
\cite{DulcicPRB03,ShibataPRB03} at a single frequency presented very
puzzling data for the vortex parameters, in particular strong field
dependence of the vortex viscosity (or, using the Bardeen-Stephen
model, for the upper critical field itself).  These findings have
their counterpart in other experimental and theoretical results: the
London penetration depth was found theoretically to strongly depend on
impurity level \cite{GolubovPRB02}, thus giving an explanation for the
large range of reported values.  $\mu$SR spectroscopy data required
for their satisfactory explanation that the application of a magnetic
field induces a strong increase in the penetration depth at low fields
\cite{OhishiJPSJ03}, or at least that two different characteristic
lengths \cite{ServentiPRL04} were involved.  Moreover, it was
theoretically shown \cite{BabaevPRL02} that the structure of vortices
is very different in two-gap, with respect to single-gap,
superconductors.  To our knowledge, there are no exhaustive theories at
present for what concerns the dynamics of vortices in multigap
superconductors.  There are however several experimental indications
that a sufficiently strong magnetic field (of order 1 T) can quench
the two-gap nature of MgB$_{2}$: scanning tunnel \cite{EskildsenPRL02}
and point-contact \cite{GonnelliPRB04} spectroscopies showed that the
superconductivity coming from the $\pi$ band is strongly suppressed
with the application of an external field $\sim$0.5 T. Neutron
spectroscopy \cite{CubittPRL03} brought evidence for a transformation
of the vortex lattice at a similar field.

From all the above considerations it appears that, for different
reasons, the microwave vortex state properties in YBCO (and, in
general, in rare earth substituted materials RE-BCO) and MgB$_{2}$
are not yet unanimously understood. Aim of this study is to present
microwave measurements at different frequencies and in wide
magnetic field ranges, in order to identify the main fundamental
mechanisms acting on the complex response. A related interest is to
determine whether simple vortex motion models can be safely applied,
possibly confining the complexity in some lumped parameter, or
different approaches are needed.

This Chapter is organized as follows: Sec.\ref{theory} presents an
overview of the mean-field theory for vortex motion and two-fluid
complex response, with some extension due to the peculiar electronic
structure of the superconductors under study.
Sec.\ref{experimental} presents the main properties of the samples
under study, the electromagnetic response in thin films, and some
detail on the resonant cavity and the Corbino disk employed for the
measurements.  Sec.\ref{results} presents and discusses the
results in YBCO, MgB$_{2}$ and SmBCO. Conclusions are drawn in
Sec.\ref{conc}.

\section{THEORETICAL BACKGROUND}
\label{theory}
Considering an electromagnetic field incident on a flat interface
between a generic medium and a bulk, thick (with respect to the
penetration depth and skin depth) (super)conductor, the response to
the field is given by the surface impedance $Z_{s}$ \cite{Jackson}.
$Z_{s}$ equals the ratio between the tangential components of the
alternating electric and magnetic fields,
$Z_{s}=\frac{E_{\parallel}}{H_{\parallel}}$.  This expression can be
easily cast in the form $Z_{s}=\mathrm{i}\omega\mu_{0}\tilde\lambda$,
being $\tilde\lambda$ an appropriate complex screening length.  In the
normal state, the screening length $\tilde\lambda\rightarrow
\frac{\delta_{n}}{\sqrt{2\mathrm{i}}}$, where
$\delta_{n}^2=2\rho_{n}/\mu_{0}\omega$ is the skin depth and $\rho_{n}$
is the normal state resistivity.  Deep in the superconducting state
($T\rightarrow0$), the screening length is the London penetration
depth $\lambda$.  The complex response can be also described by a
complex resistivity $\widetilde{\rho}$, related to the complex screening
length $\tilde\lambda$ through the relation
$\widetilde{\rho}=\mathrm{i}\omega\mu_{0}\tilde\lambda^{2}$ (or equivalently
to the surface impedance $Z_s$ through the relation $Z_s =
\sqrt{\mathrm{i}\omega\mu_0\widetilde{\rho}}$).  Thus, the response
can be expressed formally either by the surface impedance, the complex
resistivity, the complex conductivity, or the complex screening
length.

Many models have been developed and used for the frequency response in the
vortex state, with various degrees of complexity
\cite{OngPRB97,PortisEL88,SoninJETP89,MarconPRB91,CCPRL91,BrandtPRL91,
SoninPRB92,PlacaisPRB96}.  A very general expression for the
surface impedance of a semi-infinite superconductor in the mixed state
has been calculated \cite{CCPRL91}, and later extended
\cite{CCPRB92a,CCPRB92b,CCPRB93}, by Coffey and Clem (CC) within the limit
of validity of the two-fluid model (that is, in the local response
limit).  The result was expressed in terms of the combination of three
complex screening lengths, related to the different contributions to the
overall e.m. response: the superfluid response, given by the
temperature and field dependent London penetration depth
$\lambda(T,B)$; the normal fluid skin depth, given by
$\delta_{nf}(T,B,\omega) =
\left[2/\mu_0\omega\sigma_{nf}(T,B)\right]^{1/2}$ with $\sigma_{nf}$
the quasiparticle conductivity; and the vortex
response, given by the complex vortex penetration depth
$\widetilde\delta_v(T,B,\omega)=\left[2\rho_{v}(T,B,\omega)/\mu_0\omega\right]^{1/2}$.
The resulting expression for the surface impedance reads:
\begin{equation}
Z_s(T,B) = \mathrm{i}\omega\mu_0\widetilde\lambda =
\mathrm{i}\omega\mu_0\left(\frac{\lambda^2(T,B)-(\mathrm{i}/2)\widetilde\delta^2_v(B,T,\omega)}{1+2\mathrm{i}\lambda^2(T,B)/\delta^2_{nf}(B,T,\omega)}\right)^{1/2}
\end{equation}
In terms of the complex microwave resistivity of the superconductor
one writes $\widetilde\rho = \rho_1+\mathrm{i}\rho_2 =
\mathrm{i}\omega\mu_0\widetilde\lambda^2$ and, after some algebra, one
gets:
\begin{eqnarray}
\nonumber \rho_1 & = & \frac{1}{1+4(\lambda/\delta_{nf})^4}
\left[r_{1}(B,T,\omega)+2\left(\frac{\lambda}{\delta_{nf}}\right)^2r_{2}(B,T,\omega)\right]\\
\label{genform}
& &\\
\nonumber
\rho_2 & = & \frac{1}{1+4(\lambda/\delta_{nf})^4}
\left[r_{2}(B,T,\omega)-2\left(\frac{\lambda}{\delta_{nf}}\right)^2r_{1}(B,T,\omega)\right]
\end{eqnarray}
where $r_1 = \mathrm{Re}[\rho_v]$,
$r_2=\mathrm{Im}[\rho_v]+2\rho_{nf}\left(\frac{\lambda}{\delta_{nf}}\right)^2$
and $\rho_{nf}=1/\sigma_{nf}$.

Eqs.(\ref{genform}) contain in a selfconsistent way both the
quasiparticle contribution (through $\lambda$ and $\delta_{nf}$) and
the motion of vortices (through $\rho_v$).  The model is founded on
the interaction between charge carriers and a system of magnetic
vortices moving under the influence of $rf$ currents and pinning
phenomena.  Charge carriers are thought as bearing superconducting
currents, represented by a superfluid characterized by the London
penetration depth $\lambda$ which gives rise to an imaginary
conductivity $1/\mu_0\omega\lambda^2$, and normal currents represented
by $\sigma_{nf}$.  These equations should be considered as {\em master
equations}, since the actual dependence of both $\rho_1$ and $\rho_2$
as a function of temperature, magnetic field and frequency is dictated
by the functional dependence of $\lambda$, $\delta_{nf}$ and $\rho_v$
with respect to the same parameters.  In particular, $\lambda$ and
$\delta_{nf}$ (through $\sigma_{nf}$) may vary as a function of
temperature and field in different ways depending, e.g., on the order
parameter symmetry.  On the other hand, $\rho_v$ may depend on
temperature, magnetic field and frequency in several different ways,
depending on the pinning strength, inter-vortices interactions and
periodicity of the pinning potential.  This will in general result in
rather different dependencies for both the real and the imaginary part
of the resistivity.

In order to catch some general feature out of Eqs.(\ref{genform}),
we first consider the vortex motion contribution in the specific case 
of frequencies not too small, so that the long
range order of the pinning potential becomes irrelevant.  This should
be in general the case for microwave frequencies.  In this case, one
may use the specific expression of $\rho_v$ obtained for periodic pinning
potentials \cite{CCPRL91,BrandtPRL91}:
\begin{eqnarray}
\nonumber
\label{CCformulae}
\mathrm{Re}[\rho_v] & = & \rho_{ff}\frac{\epsilon+ (\nu /
\nu_0)^2}{1+(\nu/\nu_0)^2}\\
& & \\
\nonumber \mathrm{Im}[\rho_v] & = &
\rho_{ff}\frac{1-\epsilon}{1+(\nu/\nu_0)^2}\frac{\nu}{\nu_0}
\end{eqnarray}
where $\epsilon$ is a creep factor that ranges from $\epsilon=0$ (no
flux creep) to $\epsilon=1$ (free vortex motion), and $\nu_0$ is a
characteristic frequency which, in absence of creep phenomena,
corresponds to the depinning frequency $\nu_p=\kappa_{p}/2\pi\eta$.
It is worth to stress that, although Eqs.(\ref{CCformulae}) have been
obtained within a specific assumption, similar behaviors as a
function of frequency can be reasonably assumed even in more general
cases: for vortices interacting with pinning centers, in fact, it is
always possible to define a characteristic frequency $\nu_0$ which
separates a low frequency regime, in which the vortices cannot follow
the alternate Lorentz force acting on them, from a high frequency
regime where vortices oscillate in-phase around their equilibrium
positions even when strongly pinned.  If the frequency is swept across
$\nu_0$, the observed $\mathrm{Re}\left[\rho_{v}\right]$ increases
from a low frequency value to the free flow value, while
$\mathrm{Im}\left[\rho_{v}\right]$ presents a maximum at $\nu_{0}$.
This is qualitatively the behavior predicted by
Eqs.(\ref{CCformulae}).

In addition, if it is further verified that $\omega\tau_{qp}\ll$1,
$\sigma_{nf}$ is real and does not depend on frequency.  One may write
$2(\lambda/\delta_{nf})^2 = \nu/\nu_s$, with $\nu_s =
1/2\pi\mu_0\sigma_{nf}\lambda^2$.
\begin{figure}[t]
\begin{center}
\begin{minipage}[h]{80mm}
\epsfig{file=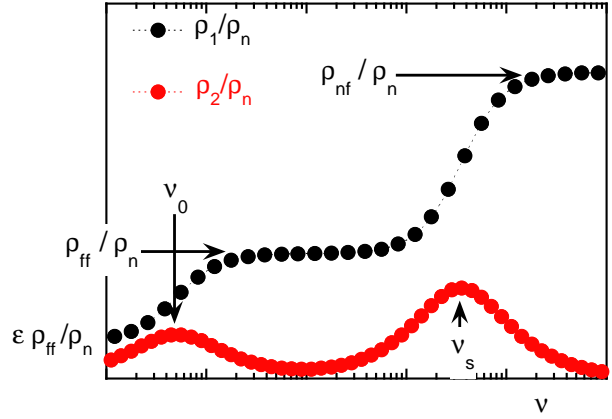, width=80mm}
\end{minipage}
\caption{Exemplification of $\rho_1(\nu)$, $\rho_2(\nu)$ curves
(frequency in a logarithmic scale) according to Eqs.(\ref{genform2}).
Also shown, the role of the main parameters.}
\label{typical_nu}
\end{center}
\end{figure}
With these substitutions
Eqs.(\ref{genform}) can be explicitly rewritten as follows:
\begin{eqnarray}
\nonumber
\rho_1 & = & \frac{1}{1+(\nu /\nu_s)^2}
\left[\rho_{ff}\frac{\epsilon+ (\nu /
\nu_0)^2}{1+(\nu/\nu_0)^2}+\frac{\nu}{\nu_s}\left(\frac{\nu}{\nu_s\sigma_{nf}}+\rho_{ff}\frac{1-\epsilon}{1+(\nu/\nu_0)^2}\frac{\nu}{\nu_0}\right)\right]\\
\label{genform2}
& &\\
\nonumber \rho_2 & = & \frac{1}{1+(\nu / \nu_s)^2}
\left[\frac{\nu}{\nu_s\sigma_{nf}}+\rho_{ff}\frac{1-\epsilon}{1+(\nu/\nu_0)^2}\frac{\nu}{\nu_0}-\frac{\nu}{\nu_s}\left(\rho_{ff}\frac{\epsilon+
(\nu / \nu_0)^2}{1+(\nu/\nu_0)^2}\right)\right]
\end{eqnarray}
It is interesting to discuss some useful limits of 
Eqs.(\ref{genform2}).

{\it The conventional low temperature limit.} At temperature low
enough, so that creep phenomena are not relevant ($\epsilon=0$ and
$\nu_0=\nu_p$) and the normal fluid conductivity can be neglected, one
has $\sigma_{nf}\rightarrow0$ and $\nu_s\rightarrow\infty$, so that
\begin{eqnarray}
\nonumber
\rho_1 & = &
\rho_{ff}\frac{1}{1+(\nu_p/\nu)^2}\\
\label{gittl-ros}
& &\\
\nonumber
\rho_2& = &
\mu_0\omega\lambda^2+\rho_{ff}\frac{\nu_p/\nu}{1+(\nu_p/\nu)^2}
\end{eqnarray}
which are equivalent to the conventional Gittleman-Rosenblum
expressions for the microwave resistivity in the mixed state,
Eqs.(\ref{semplice}), apart from the first term in the imaginary part,
representing the zero-field imaginary conductivity.

{\it The two-fluid limit.} As a second example, it is easily seen
that for $B\rightarrow 0$ one has $\rho_{ff}=0$ and
Eqs.(\ref{genform2}) reduce to the two-fluid conductivity,
Eqs.(\ref{twofluid}).

{\it The high frequency limit: free flow.} When the measuring
frequency is much larger that the vortex characteristic frequency,
$\nu\gg\nu_0$, Eq.(\ref{genform2}) reduces to
\begin{eqnarray}
\nonumber
\rho_1 & \simeq & \frac{1}{1+(\nu /\nu_s)^2}
\left[\rho_{ff}+\frac{1}{\sigma_{nf}}\left(\frac{\nu}{\nu_s}\right)^2\right]\\
\label{genform_hf}
& &\\
\nonumber
\rho_2 & \simeq & \frac{\nu/\nu_s}{1+(\nu / \nu_s)^2}
\left(\frac{1}{\sigma_{nf}}-\rho_{ff}\right)
\end{eqnarray}
In this limit vortices oscillate in phase making very short
displacements from their equilibrium position.  Despite the possibly
finite pinning, as could be determined by, e.g., dc resistivity or
magnetization, the vortex response coincides with the free flux flow.
In this case the response depends on intrinsic physical quantities
only: $\nu_s$, $\rho_{ff}$ and $\sigma_{nf}$.  The analysis of the
data in this limit is more stringent, since only three parameters are
involved (and, in addition, $\nu_s$ and $\sigma_{nf}$ are
related one to each other).  It should be noted that this limit might
apply to various measurements at the high edge of the microwave
spectrum.

Up to now we have mainly considered the vortex motion.  However, both
superfluid and quasiparticle densities are affected by the magnetic
field.  In fully gapped superconductors, such as conventional
superconductors, the quasiparticle density of states (DOS) at Fermi
level is nonzero only within the vortex core.  As a consequence, the
total DOS is proportional to the area occupied by the cores and the
field-dependent depletion of the superfluid fractional density is
simply $\Delta x_{s}\propto\frac{\xi^{2}}{R^{2}}\approx B/B_{c2}$,
where the coherence length $\xi$ gives the dimension of the vortex and
$R\sim\sqrt{\Phi/B}$ is the intervortex spacing, with $\Phi$ the total
magnetic flux through the sample.

The situation is rather different in superconductors with lines of
nodes in the gap.  In this case, it has been shown that extended
states {\it outside} of the vortex core have the most relevant weight
\cite{VolovikJETP93,WonPRB96}.  However, such states have zero DOS
only at the Fermi level, and any finite energy brings a variation of
the quasiparticle fractional density.  As an example, circulating
supercurrents around vortices give rise to a Doppler shift
\cite{YipPRL92,VolovikJETP93} of the quasiparticle energy.  This
results in a depletion of the superfluid fractional density which is
proportional to the spatial range of the circulating
supercurrents, $min\{\lambda,R\}\approx R$ for fields not too low, and
to the number of vortices $\sim B/\Phi$.  As a result, one has $\Delta
x_{s}\approx\sqrt{B/B_{pb}}$ where $B_{pb}$ is a characteristic
pair-breaking field, that can depend on the gap gradient at the nodes,
on impurity scattering and temperature, but it can be assumed
(as a first approximation) to be $B_{pb}\approx c B_{c2}$ \cite{WonPRB96},
with $c\sim o(1)$.

Thus, in general one can write:
\begin{equation}
      x_{s}(T,B)=x_{s0}(T)-\Delta x_{s}(T,B) \approx
      x_{s0}(T)\left[1-\left(\frac{B}{cB_{c2}}\right)^{\alpha}\right]
      \label{nodi}
\end{equation}
where $\alpha$ depends on the symmetry of the gap and on impurities
\cite{NakaiPRB04}, and in clean cases
$\alpha =$ 1 or $\alpha =\frac{1}{2}$ in fully gapped superconductors
and in superconductors with lines of nodes, respectively. More
accurate treatments \cite{DahmPRB02,LaihoPRB04} do not change the
qualitative result of a significant reduction of the superfluid density with
the application of a magnetic field.

It is useful to plot some illustrative curve of the theoretical
predictions, in order to elucidate the role of the different
parameters.  We will also make use of the limits above discussed.

We present first schematic shapes of the curves $\rho_1(\nu)$ and
$\rho_2(\nu)$ as obtained from the general expression,
Eqs.(\ref{genform2}).  We stress that when plotting resistivity vs.
frequency it is not necessary to assume any specific temperature or
field dependence for the quantities appearing in the equations.  As
shown in Fig.(\ref{typical_nu}), the shape of the two curves
$\rho_1(\nu)$ and $\rho_2(\nu)$ is affected in different ways by the
various parameters ($\rho_{ff}, \epsilon$ and $\nu_0$ determine the
vortex motion, $\sigma_{nf}$ and $\nu_s$ the properties of the
quasiparticles).  In particular, it is seen that even a rather simple
model can give significantly different curves for $\rho_1(\nu)$ and
$\rho_2(\nu)$.  Since the variation range of the involved parameters
is very wide\footnote{All these parameters are expected to vary as a
function of temperature and magnetic field.  In particular,
$\nu_s\rightarrow 0$ and $\sigma_{nf}\rightarrow \sigma_n$ as the
superconductor enters the normal state by increasing the field or
temperature; for fields not too low, $\rho_{ff} \sim B$; $\epsilon
\sim 0$ at low temperatures and $\epsilon \rightarrow 1$ at higher
$T$.  Finally, the behavior of $\nu_0$ might be extremely dependent
on the pinning characteristics of the specific sample.}, it is evident
that a wide measuring frequency range can prove especially useful in
the interpretation of the data.

In this Chapter we will report also data for the complex resistivity
at a single frequency (resonant cavity technique) and as a function of
the field, so that it is appropriate to present and discuss some
theoretical curve at a fixed frequency.  In order to plot the
microwave resistivity {\it vs.} the magnetic field, it is necessary to
make some additional hypotheses with respect to the frequency plots.
In particular, it is necessary to assume the field dependence of the
parameters defining the frequency dependence of the resistivity.  In
order to limit the number of the assumptions, we restrict ourselves to
the case of sufficiently high frequency ($\nu\gg\nu_0$) so that the
field variations of $\nu_0$ (and, to a larger extent, of $\epsilon$)
become almost irrelevant.  We also assume that the vortex viscosity
$\eta$ does not depend on field.  For the quasiparticle and superfluid
response, we assume that $\sigma_{nf} = x_n(T,B)\sigma_n$ and
$\lambda^2 = \lambda_0^2/x_s(T,B)$, with $x_n$ being the normal
fractional density, related to $x_s$ through the relation $x_n+x_s=1$.
We momentarily neglect in this illustrative case possible differences
between the scattering times of quasiparticles and normal electrons.
\begin{figure}[t]
\begin{center}
\begin{minipage}[h]{120mm}
\epsfig{file=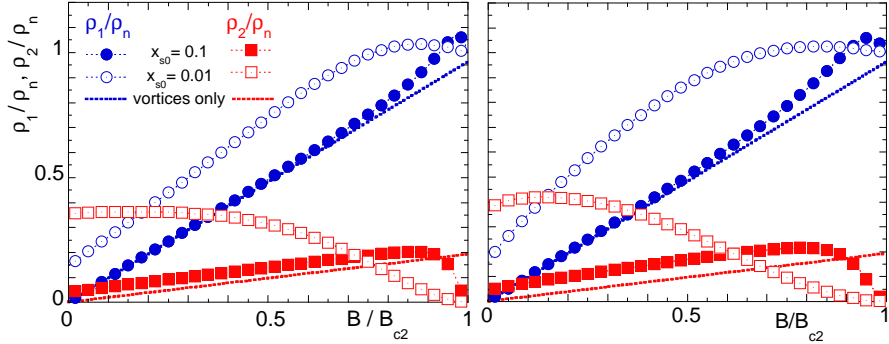, width=120mm}
\end{minipage}
\caption{Reduced field dependence of the vortex state complex
resistivity according to the CC model \cite{CCPRL91},
Eq.(\ref{genform}), for $\nu/\nu_{0}=5$.  The curves for small (full
symbols) and very small (open symbols) superfluid concentration are
reported, and compared to the vortex motion contribution alone
($\mathrm{Re}[\rho_v]$, $\mathrm{Im}[\rho_v]$ as in
Eq.(\ref{CCformulae}), dashed lines).  Parameters are reported in the
text.  Left panel, fully gapped superconductors.  Right panel,
superconductor with lines of nodes in the gap (see text).}
\label{sd-like}
\end{center}
\end{figure}
\begin{figure}[t]
\begin{center}
\begin{minipage}[h]{80mm}
\epsfig{file=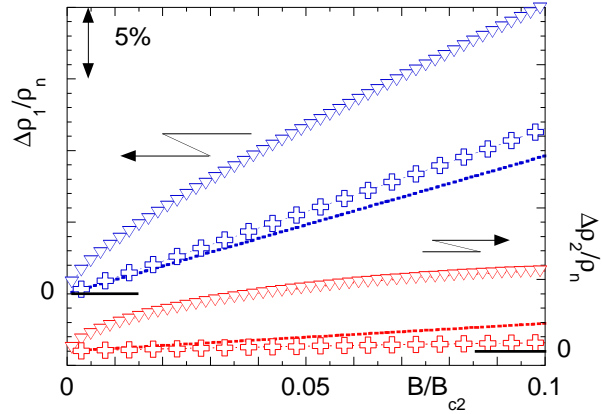, width=80mm}
\end{minipage}
\caption{Plot of the field variation of the complex resistivity, at
low reduced fields.  Parameters as in Fig.\ref{sd-like}, with
$x_{s0}=$ 0.01.  Blue symbols, left scale:
$\Delta\rho_{1}/\rho_{n}=\left[\rho_{1}(B)-\rho_{1}(0)\right]/\rho_{n}$.
Red symbols, right scale:
$\Delta\rho_{2}/\rho_{n}=\left[\rho_{2}(B)-\rho_{2}(0)\right]/\rho_{n}$.
Dotted lines, vortex motion only.  Crosses, fully gapped
superconductor.  Triangles, superconductor with lines of nodes in the
gap.  A sublinear field dependence is clear when the superconducting
gap has lines of nodes.}
\label{smallB}
\end{center}
\end{figure}
We plot the calculated curves at the fixed frequency $\nu=50$ GHz as a
function of the reduced applied magnetic field in two cases: fully
gapped superconductor and superconductor with lines of nodes in the
gap, in the left and right panels of Fig.\ref{sd-like}, respectively.  Curves
are plotted for different values of the zero field superfluid density
$x_{s0}$ (which is equivalent to different temperatures).  We use
typical values for high $T_c$ superconductors, $\lambda_0$ = 1000
\textrm{\AA} and $\sigma_{n}=10^6$ S\textperiodcentered m$^{-1}$.  We
also fix $\nu_0=10$ GHz as reported for HTCS at low temperature
\cite{WuPRL95,TsuchiyaPRB01,SartiPhC04} and $\epsilon=0$, both
independent on field.
In both cases at relatively large superfluid concentration
($x_{s0}=0.1$) the real part of the resistivity is substantially
coincident with $\mathrm{Re}[\rho_v]$ at low fields ($\rho_v$ is the
resistivity due to the vortex motion, and it does not depend, by
definition, on the superfluid density $x_s$).  This is due to the fact
that with the values used for $\lambda_0$ and $\sigma_n$, $\nu_s \gg
50$ GHz at low temperatures and fields, so that, as discussed
previously, Eqs.(\ref{genform2}) reduce to the Gittleman-Rosenblum
expression, Eq.(\ref{semplice}).  However, this is no longer the
case at very small superfluid density (i.e., at higher temperatures
close to $T_c$ or with strong pair breaking), $x_{s0}=0.01$, where
neglecting the quasiparticle contribution results in a rather large
error.  Interestingly, as reported in Fig.\ref{smallB}, in the case of
lines of nodes in the superconducting gap, at low fields the variation
of the imaginary part of the resistivity is clearly found to increase
as $\sim\sqrt{B}$ even in presence of flux flow (linear in the applied
field), while for fully gapped superconductors this increase turns out
to be linear with $B$.  Moreover, a closer inspection of the
$\rho_1(B)$ curve shows that a partial $\sqrt{B}$ dependence is also
present in the real part of the resistivity, for superconductors with
lines of nodes in the gap, at sufficiently low superfluid
concentration.
It is important to stress that the role of quasiparticles is relevant
whenever $\nu_s$ is not much larger than the measuring frequency
$\nu$.  Due to the parameters used, in the case above exemplified
$\nu_s$ is always much larger that $\nu$ (50 GHz) unless
$x_s\rightarrow 0$, that is for $T\simeq T_c$ (or $B \simeq
B_{c2}$).

Before commenting experimental results concerning different
superconductors, we now briefly discuss which are the main possible
differences between the relatively simple scenario depicted up to now
and the real world.  First of all, we have neglected any creep
phenomena for vortices ($\epsilon=0$), which is unlikely to be correct
at high temperatures.  A non-vanishing $\epsilon$ will strongly
influence the low frequency ($\nu \lesssim \nu_0$) resistivity.  More
important, since $\nu_0=\kappa_{p}/2\pi\eta$ only for $\epsilon=0$, an
increasing $\epsilon$ could introduce rather strong temperature and/or
field dependencies in $\nu_0$.  In particular, since $\nu_0$ is
expected to grow rather fast with increasing $\epsilon$, in a
measurement at fixed frequency one might have $\nu>\nu_0$ at low
temperatures or fields and $\nu<\nu_0$ at higher temperatures or
fields.  This will result in unpredictable shapes for the curves
$\rho(H)$ or $\rho(T)$, depending on the field/temperature variation of $\nu_0$.

The second main hypothesis which is not necessarily verified concerns
the quasiparticle conductivity.  Smaller values of $\nu_{s}$ can
substantially change the quasiparticle contribution to the
resistivity.  As an example, quasiparticle relaxation rates shorter
than the normal state values, as reported by several authors
\cite{BonnPRB93,NagashimaFujita94,JacobsJPCS95}, would tend to
decrease substantially $\nu_{s}$.  One might have as a consequence
large quasiparticle contributions even at low temperatures and
fields.

Finally, it must be noticed that Eqs.\ref{CCformulae} for the
vortex resistivity are obtained in the rather special case of periodic
pinning potential. Moreover, they do not take into account the
resistivity due to Josephson or Abrikosov-Josephson vortices, whose
dynamics might be rather different with respect to standard vortices.
The expression for the vortex resistivity $\rho_v$ might then be rather
different, in the most general case.  However, we will show in the
following that most of the main features observed in the microwave
data of several superconducting materials can be quantitatively
explained using the theoretical expressions cited up to now.  We will
come back to specific points during the discussion of the data in
Sec.\ref{results}
\section{EXPERIMENTAL SECTION}
\label{experimental}
\subsection{Samples}
\label{samples}
All measurements here presented were performed in thin, high-quality
superconducting films.  YBa$_{2}$Cu$_{3}$O$_{7-\delta}$,
SmBa$_{2}$Cu$_{3}$O$_{7-\delta}$, and MgB$_{2}$ were investigated
extensively.  Samples were squares, of side $l$ and thickness $d$.
Substrates have been carefully chosen for microwave measurements.  The
crystal structure was investigated by X-ray $\Theta-2\Theta$
diffraction.  The $c$-axis orientation was assessed by measuring the
full-width-half-maximum (FWHM) of the rocking curve of an appropriate
peak.  In-plane X-ray $\Phi-$scan was measured in samples Y1, Y2, Y3,
Sm1, Sm2, and M1, showing excellent in-plane epitaxiality.  Surface roughness
was investigated by AFM over typical 1 $\mu$m $\times$ 1 $\mu$m area.
All cuprates sample were nearly optimally doped or slightly overdoped.
$T_{c}$ and the resistivity above $T_{c}$, $\rho_{0}$, were estimated
by electrical transport methods.  Depending on the sample, we used
$dc$ or microwave resistivity.  In this latter case $T_{c}$ was
estimated from the inflection point of the temperature-dependent real
part of the microwave resistivity (this temperature is found to
coincide with the temperature where the real and imaginary parts of
the microwave fluctuational conductivity equal one to the other, which
is an accurate evaluation of the mean-field critical temperature
\cite{SilvaEPJB04}), and $\rho_{0}$ from the measured real part
$\rho_{1}$ in zero magnetic field.  Typical $\pm$0.5 K uncertainties
of these methods are inessential for the purposes of the present
paper.  Material parameters and appropriate references are reported in
Table \ref{tabella}.  More details on sample preparation and
characterization are reported in the References (see Table).
\begin{table}
{\tiny \begin{tabular}{@{}llllllll}
Material & YBCO & YBCO & YBCO & YBCO & SmBCO & SmBCO & MgB$_{2}$\\
Sample no. & Y1 & Y2 & Y3 & Y4 & Sm1 & Sm2 & M1\\
Substrate & LaAlO$_{3}$ & LaAlO$_{3}$ & LaAlO$_{3}$ & CeO$_{2}$/YSZ &
LaAlO$_{3}$ & LaAlO$_{3}$ & sapphire\\
Thickness (nm) & 220 & 220 & 220 & 200 & 220 & 220 & 100\\
Lateral dimension (mm) & 10 & 10 & 10 & 10 & 10 & 10 & 5\\
$\Theta-2\Theta$ FWHM & 0.1$^{\circ}$ & 0.1$^{\circ}$ & 0.1$^{\circ}$ &
0.2$^{\circ}$ & 0.2$^{\circ}$ & 0.2$^{\circ}$ & N.A.\\
Surface roughness (nm) & 2 & 2 & 2 & N.A. & 3 & 3 & N.A.\\
$T_{c}$ (K) & 89.5 & 89.5 & 90 & 88 & 87 & 87 & 36\\
$\rho_{0}$ ($\mu\Omega\cdot$cm) & 130 & 140 & 130 & 250 & 300
& 300 & 5\\
References & \cite{BeneduceIJMPB99,NeriPhC00} &
\cite{BeneduceIJMPB99,NeriPhC00} &
\cite{BeneduceIJMPB99,NeriPhC00} & \cite{CamerlingoINFM02} &
\cite{BoffaIJMPB03,BoffaPhC03} & \cite{BoffaIJMPB03,BoffaPhC03} &
\cite{FerrandoSUST03}\\
\end{tabular}
}
\caption{Data for the structural and electrical characterization of the
samples investigated.  $\rho_{0}=\rho_{1}$(100 K, 48 GHz) and
$\rho_{dc}$(40 K) in RE-BCO and MgB$_{2}$, respectively.  N.A.: not
available.}
\label{tabella}
\end{table}
\subsection{Microwave response in thin films.}
\label{thin}

Since in all the measured samples the thickness is of the order of, or
smaller than, the commonly reported values for the penetration depth,
it is appropriate to shortly describe the electromagnetic response of
an electromagnetically thin (super)conducting film.

In the case of bulk samples, for which the sample thickness is much
greater than the electromagnetic field penetration depth (of the order
of $\mathrm{min}(\lambda, \delta_{nf})$), the surface impedance is
given by the usual expression already mentioned: $Z_{s}=\left ({\rm
i}\omega\mu_{0}\tilde{\rho}\right )^{1/2}$.

When the sample thickness $d\lesssim \mathrm{min}(\lambda,
\delta_{nf})$, the electromagnetic field is transmitted through the
film and reaches the substrate and any supporting layer, usually a
metallic backplate.  The field is therefore determined by the
interaction with both the underlying layers and the film of finite
thickness.  In this situation the simple expression for $Z_{s}$ no
longer holds, being substituted by an effective surface impedance
$Z^{eff}_{s}$ which can be derived by means of standard impedance
transformation relations \cite{Collin}:
\begin{equation}
\label{zeff}
Z^{eff}_{s}=Z_{s}
\frac{Z^{eff}_{d}+{\rm i} Z_{s}\tan(k_{s}d)}{Z_{s}+{\rm i}
Z^{eff}_{d}\tan(k_{s}d)}
\end{equation}
where $k_{s}=\mu_{0}\omega/Z_{s}$ is the HTCS propagation constant and
$Z^{eff}_{d}$ is the effective surface impedance of the substrate.
The full expression of Eq.(\ref{zeff}) can be significantly simplified
when two main conditions are met: $|Z^{eff}_{d}|\gg|Z_{s}|$, meaning
that the substrate contribution can be neglected, and $d\ll
\mathrm{min}(\lambda, \delta)$, which is usually the case for epitaxially
grown high $T_{c}$ superconductor films.  In this case the so-called
thin-film approximation \cite{SridharJAP88,SilvaSUST96} is obtained:
\begin{equation}
\label{zf}
Z^{eff}_{s}\simeq\frac{\tilde{\rho}}{d}
\end{equation}
The applicability of this equation heavily depends on the (possibly
temperature-dependent) properties of the substrate.  Metallic
\cite{BeeliPhC00} and semiconducting \cite{PompeoEU03,PompeoSUST05}
substrates strongly affect $Z^{eff}_{s}$ and they impose the use of
the full expression, Eq.(\ref{zeff}).  On the other hand dielectric
substrates with backing metallic plate have often impedances high
enough so that Eq.(\ref{zf}) can safely be used\footnote{Noticeable
exceptions are dielectrics with strong temperature dependent
permittivity \cite{SilvaSUST96,KleinJAP90,HartemannIEEE92,HeinJAP94}
or with an accidentally unfavorable combination of thickness,
permittivity and operating frequency \cite{SilvaPhC97}.}.
\subsection{Cavity measurements.}
\label{cavity}
The microwave response at high frequency was measured in RE-BCO by the
end-wall cavity technique \cite{PooleESR} at 48.2 GHz.  The
cylindrical cavity, of 8.2 mm diameter, was inserted in a liquid/solid
Nitrogen cryostat, so that temperatures in the range 60-150 K could be
reached.  Cryogenic waveguides were used in order to couple the
one-port cavity to the external microwave source.  The temperature of
the entire cavity, including the sample, could be stabilized for hours
within $\pm$10 mK. A magnetic field $H$ was applied along the $c$ axis
and supplied by a conventional electromagnet.  The maximum attainable
field in this setup was $\mu_{0}H\leq$0.8 T. In the following, we will
assume that inside the sample $B \simeq\mu_0 H$ \footnote{We note that
demagnetization effects determine a penetration field much smaller
than $H_{c1}$ in thin films \cite{ClemPRB94}, and field inhomogeneities inside
the sample are expected (and also directly found \cite{XingJAP94}) to be
irrelevant in thin films for fields greater than a few mT in our
temperature range.}.

\begin{figure}[t]
\begin{center}
\begin{minipage}[h]{80mm}
\epsfig{file=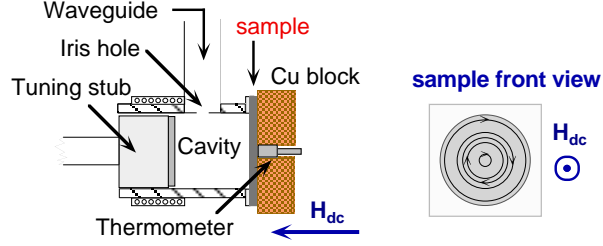, width=80mm}
\end{minipage}
\caption{Left: straight section of the resonant cavity.  Right:
current patterns on the square sample.}
\label{figcavity}
\end{center}
\end{figure}

For the measurements, undercoupling of the cavity and the TE$_{011}$
resonant mode were chosen.  In this mode the currents flow along
circular paths on the end-wall occupied by the sample, have zeroes at
the center of the end walls and at the joints between the end walls
and the body of the cavity, and maxima at half the radius of the end
wall.  This configuration allows to neglect the losses due to
imperfect electric contact between the sample and the cavity;
moreover, it makes possible the use of a mechanical stub to tune the
cavity at the desired resonant frequency (in our case, in the range
48.0 $\pm$ 0.5 GHz).  The degenerate TM mode, strongly suppressed by
electrical isolation between the end walls and the body of the cavity,
was further shifted in frequency by appropriate mode traps.
Fig.\ref{figcavity} reports a sketch of the cavity setup and of the
pattern of the currents on a typical 10 mm $\times$ 10 mm sample.

Measurements of the field and/or temperature induced changes of the
quality factor $Q$ and resonance frequency $f_{0}$ yielded changes in
the effective surface impedance through the well known expression
$\Delta Z_{s}^{eff}(H,T)=G\Delta \left[\frac{1}{Q(H,T)}-2\mathrm{i}
f_{0}(H,T)\right]$, where $G$ is an appropriate geometrical
factor \cite{Collin}. Separate calibration of the response of the cavity
allowed for the determination of the absolute value of $R_{s}^{eff}$.
In most of the measurements here reported, $Q$ and
$f_{0}$ were measured as a function of the magnetic field at
various fixed temperatures in the range 60 K - $T_{c}$.  Measurements
were performed by sweeping the field from zero up to 0.8 T either
after zero-field-cooling (ZFC) to the desired temperature, or by
increasing the temperature after each field sweep. We did not observe
hysteresis, apart a 10\% effect at temperatures below 70 K in some of
the samples. In some cases, to check the relevance of the hysteresis,
full magnetic cycles were performed. Measurements here reported refer
only to the cases where hysteresis is absent, or well below 10\% of the
total response.

From the field induced change of $Q$ and $f_{0}$ the field induced
change of the effective surface impedance $\Delta Z_{s}^{eff} =
Z_{s}^{eff}(H,T)-Z_{s}^{eff}(0,T)$ could be obtained
\cite{SilvaMST98}.  Due to the small thickness of the films, one has
$Z_{s}^{eff}=\tilde\rho/d=\rho_{1}/d+{\mathrm{i}} \rho_{2}/d$, where
$\tilde\rho$ is the complex resistivity.

The main experimental errors are due to the following reasons: the
absolute value of $Z_{s}^{eff}$ is affected by errors in the
calibration of the cavity and in the evaluation of the geometrical
factors \cite{CeremugaJS95,MazierskaJS97,Silvareview}.  Additionally,
$\tilde\rho$ has an intrinsic uncertainty due to the evaluation of the
film thickness (typically 10\%).  By contrast, calibration of the
cavity does not affect the field variation of $\tilde\rho$ at fixed
temperature, and geometrical factors and film thickness give only a
possible overall scale factor.  All these sources of errors are
strongly reduced (if not eliminated at all) when working with reduced
complex resistivity changes, $\Delta
\tilde\rho/\rho_{0}=\Delta\rho_{1}/\rho_{0}+\mathrm{i}
\Delta\rho_{2}/\rho_{0}$, with $\rho_{0}=\rho$(100 K) in YBCO and
SmBCO. In the following the data will be reported in one of the
mentioned formats.

Finally, some useful features of the present setup should be stressed.
First, the microwave currents and fields lie in the $(a,b)$ planes,
thus avoiding any $c$ axis contribution to the measured response.
Second, the dc magnetic field is perpendicular to the microwave
currents (maximum Lorentz force configuration).  Third, the microwave
currents essentially probe only an annular region of 4.1 mm mean
diameter and $\sim$ 2 mm width, centered on the sample.  For the
subsequent discussion, it is relevant that flux lines are not forced
to cross the border of the sample: the oscillation induced by the
microwave currents are limited to the annular region above mentioned.
In particular, no edge effects are relevant in our configuration.
\subsection{Corbino disk measurements.}
\label{corbino}
\begin{figure}[t]
\begin{center}
\begin{minipage}[h]{80mm}
\epsfig{file=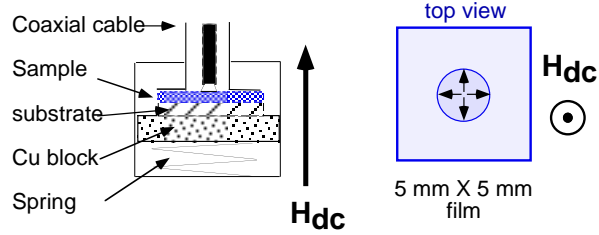, width=80mm}
\end{minipage}
\caption{Left: straight section of the Corbino disk cell.  Right:
current patterns on the square sample.}
\label{figcorbino}
\end{center}
\end{figure}

The frequency dependence of the vortex-state microwave response was
measured in YBCO and MgB$_{2}$ films.  Measurements are obtained
through a Corbino disk geometry \cite{WuPRL95}: a swept frequency
microwave radiation is generated by a vector network analyzer (VNA)
and guided to the sample under study through a coaxial cable.  The
sample, placed inside the cryomagnetic apparatus, shortcircuits the
coaxial cable.  A double spring is used to obtain a good
electrical contact between the end cable connector and the sample.  To
further improve the contact, a thin annular indium foil is placed
between the sample and the external part of the coaxial cable.  Due to
the limitations discussed below, data are reported for the frequency
range 2-20 GHz, well inside the capabilities of the VNA (45 MHz-50
GHz) and of the cables (cutoff frequency 50 GHz).  A magnetic field up
to 10 T was applied along the $c$ axis.  The temperature ranges 70 K -
100 K and 5 K - 40 K were explored, in YBCO
and MgB$_{2}$, respectively.  In Fig.\ref{figcorbino} we show a
sketch of the Corbino disk cell and of the current pattern on the
sample.

In principle, a measurement of the complex reflection coefficient $\Gamma_0$ 
at the sample surface directly yields the complex
effective surface impedance $Z_{eff}$ of the sample by the standard
relation
\[
\Gamma_0 = \frac{Z_{s}^{eff}-Z_0}{Z_{s}^{eff}+Z_0}
\]
where $Z_0$ is the impedance of the cable \cite{Collin}.  In our case,
however, the measured quantity is the reflection coefficient at the
instrument location $\Gamma_m$, which contains, besides the response
of the sample, reflections and attenuation due to the cable line
between the sample and the VNA. Due to the long line, necessary to
place the VNA far from the stray field of the magnet, the contribution
of the sample to the overall $\Gamma_m$ is rather small.  Further, the
response of the part of the line inside the cryostat depends on the
temperature profile across the cable, that varies in general during the
measurement.  Thus, full calibration of the cable and, consequently,
direct measurements of absolute values of the impedance of the sample
are not feasible.  To overcome this problem, we developed a custom
calibration procedure, through which we could obtain the variations of
the effective surface impedance with the temperature or with the
field, $\Delta Z_{s}^{eff}(\nu,H,T)$.  The description of the
calibration procedure is rather cumbersome, and it has been
extensively described elsewhere \cite{SartiCM04,AmabileEU03}.  Here we
recall only the major hypotheses necessary to obtain the impedance data
from the measured complex reflection coefficient.

Reliable measurements of the {\it temperature} variations of the
effective surface impedance, $\Delta
Z_{s}^{eff}(\nu,H,T)=Z_{s}^{eff}(\nu,H,T)-Z_{s}^{eff}(\nu,H,T_{ref})$,
require {\it (i)} that the temperature variation of the complex
reflection coefficient are not dominated by the change in the response
of the cable, and {\it (ii)} that at least at sufficiently low
temperature $\Gamma_0(\nu)\simeq-1$ (or, equivalently, $R_{s}^{eff}, X_{s}^{eff}\ll
Z_0$) \cite{SartiCM04}.
In particular, variations of $R_s^{eff}$ mainly reflect on variations
in the modulus of $\Gamma_0$, while variations of $X_s^{eff}$ have as
a main effect a phase change of $\Gamma_0$.  The accuracy and
reliability of the measurement of $\Delta R_s^{eff}$ and $\Delta
X_s^{eff}$ depend then on relative variations of modulus and phase of
$\Gamma_0$ with respect to corresponding variations (due to, e.g., the 
change of the cable characteristics) of $\Gamma_m$,
respectively.
It turns out that for the measurements of $R_{s}^{eff}$ the
requirements are fulfilled in the temperature ranges 5 K - 40 K and 70
K - 100 K in MgB$_{2}$ and YBCO, respectively.  Unfortunately, the
temperature variations of the phase signal due to the sample are of
the same order of magnitude of the temperature variations of the phase
due to the cable, so that a reliable measurement of $X_s^{eff}$ is not
feasible with this kind of measurement.  No such problems arise at
fixed temperature and with sweeping field: in this case, the response
of the cable is almost exactly constant (we checked that the line has
a very small magnetic response): {\it field} variations $\Delta
Z_{s}^{eff}(\nu,H,T_{0})=Z_{s}^{eff}(\nu,H,T_{0})-Z_{s}^{eff}(\nu,0,T_{0})$
can be reliably obtained.

These measurements can be converted to absolute values of $R_s^{eff}$
and/or $X_s^{eff}$ by assuming some known $R_s^{eff}(\nu)$
or $X_s^{eff}(\nu)$ at some specific temperature or field.  In general, one
can take $R_s^{eff}(\nu)\simeq 0$ at sufficiently low temperatures and
fields, while $X_s^{eff}$ can be obtained from normal state
measurements (at high enough temperatures or fields).

There are a few additional remark on the different behavior of the
two materials investigated with the Corbino disk.  First, the quality
of the electrical contact between the coaxial connector and the film
is much better in MgB$_{2}$ than in YBCO. This results in reliable
measurements in the ranges 2 - 20 GHz and 5 - 20 Ghz in MgB$_{2}$ and
YBCO, respectively.  Second, with the available magnetic field it was
possible to reach the normal state in MgB$_{2}$, so that absolute
measurements of $R_s^{eff}$ and $X_s^{eff}$ can be obtained by
increasing the field above the upper critical field.  In YBCO
measurements were taken with field-cooling the sample (thus lowering
the temperature), and only the real part of $Z_s^{eff}$ could be
obtained.  Finally, even after the calibration procedure, there still
remain detectable oscillations in the frequency-sweeps.  Those
oscillations, together with uncertainties in geometrical factors, are
eliminated by reporting the data normalized to the values measured
above $T_{c}$.

As for the cavity measurements, the thin film approximation holds
\cite{SilvaSUST96} and one has $Z_{eff} = \widetilde\rho/d$.  Most
important, like and more than in the cavity measurements, in the
Corbino disk geometry there are no contributions at all from vortex
entry or exit: the area probed by the microwave currents is a small
circle, of radius $\sim$2 mm, at the center of the sample, and
currents flow along linear paths between the inner contact and the
outer conductor, so that vortices are forced to oscillate along
circular orbits.  Again, no edge effects affect our measurements.
\section{RESULTS}
\label{results}
\begin{figure}[t]
\begin{center}
\begin{minipage}[h]{80mm}
\epsfig{file=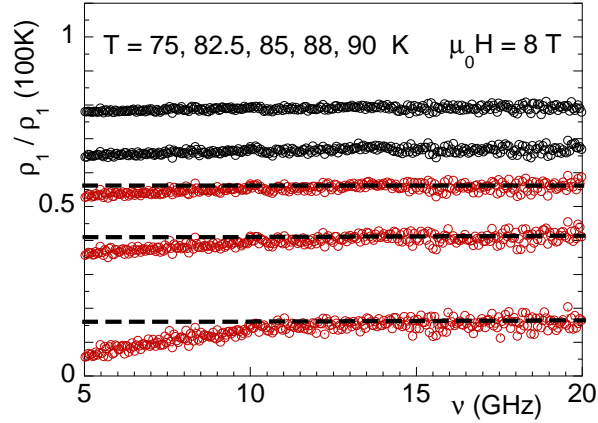, width=80mm}
\end{minipage}
\caption{Normalized real resistivity in YBCO vs. frequency at
various temperatures and $\mu_{o}H=$ 8 T. A detectable frequency
dependence only appears for $T<$85 K (red data curves). Temperature 
increases from bottom to top.}
\label{YCorb-r-nu}
\end{center}
\end{figure}
In this Section, we present and discuss the results obtained on the
various superconducting materials under study. We present separately
the results for the different materials.
\subsection{YBa$_{2}$Cu$_{3}$O$_{7-\delta}$}
YBa$_{2}$Cu$_{3}$O$_{7-\delta}$ is a somewhat paradigmatic case: as it
will be shown in the following, the microwave properties in the vortex
state follow quite closely the simple models summarized in
Sec.\ref{theory}, allowing for a rather detailed discussion of the
various dynamical regimes as a function of frequency, temperature and
magnetic field.  By means of combined wideband (5-20 GHz) and
high-frequency (48 GHz) measurements we will show (a) that it exists a
relatively wide temperature range below the critical temperature $T_c$
in which the resistivity, while clearly lower than the normal state
value, is at the same time independent on frequency, indicating that
the effect of pinning (if any) is not relevant in this $T$ range, (b)
that vortex motion follows closely the predictions of the mean-field,
Coffey-Clem theory, and (c) that approaching the critical temperature
the field dependence of the microwave resistivity points to a relevant
role of the field-dependent superfluid depletion.  We also obtain the
temperature dependence of the vortex parameters $\eta$ and
$\nu_{0}$.

\begin{figure}[t]
\begin{center}
\begin{minipage}[h]{80mm}
\epsfig{file=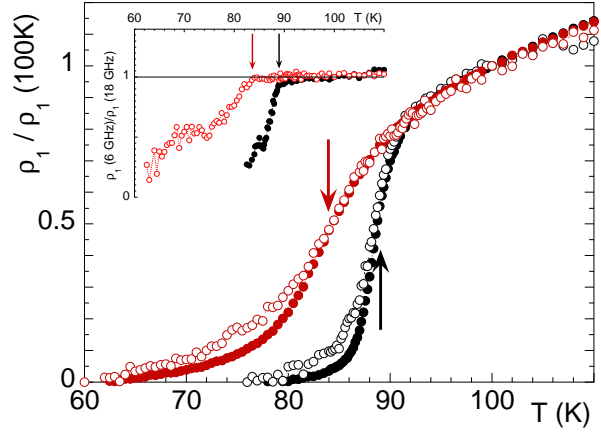, width=80mm}
\end{minipage}
\caption{Normalized real resistivity in YBCO vs.  temperature at 6 GHz
(full symbols) and 20 GHz (open symbols) and $\mu_{o}H=$ 2 T (black
symbols) and 8 T (red symbols).  It is apparent that at each field
there is no frequency dependence down to a typical temperature
$T_{f}<T_{c}$, depicted by the arrows.  The same feature is more
evident in the inset, where the ratios $\rho_1(\mathrm{6}\,
\mathrm{GHz},H,T)/\rho_1(\mathrm{20}\,
\mathrm{GHz},H,T)$ have a clear departure from 1
at $T_{f}(H)$.}
\label{YCorb-r-T}
\end{center}
\end{figure}

We begin with the frequency dependence of the microwave real
resistivity $\rho_{1}$ measured in sample Y3.  In Fig.\ref{YCorb-r-nu}
we report the frequency dependence of $\rho_{1}/\rho_{0}$ measured
with the Corbino disk at various temperatures and at a fixed field
$\mu_{0}H$= 8 T. As it can be seen, while $\rho_{1}$ is sensibly lower
than $\rho_{0}$ for $T< 90$ K, a frequency dependence is observed only
below $T = 85$ K$< T_{c}$.  This is a first indication that in the
superconducting state there is a region of the $(H,T)$ phase diagram
where the microwave response in the frequency region here investigated
is independent (or very weakly dependent) on frequency.  This finding
restricts the range of physical phenomena that can give rise to the
measured response.  More insight can be gained by representing the
data at fixed frequency as a function of the temperature, as depicted
in Fig.\ref{YCorb-r-T}, where we compare the normalized real
resistivity at two fixed frequencies (6 and 20 GHz) as a function of
temperature, for $\mu_{0}H =$ 2 T and 8 T. At each field the data at
two different frequencies lie on a single curve in a temperature range
extending from above to below $T_{c}$, while they are markedly
different at lower temperatures, below an easily estimated crossover
temperature $T_f (H)$.  This abrupt change in the frequency dependence
across $T_f$ marks the boundary between two different dynamic regimes.

The nature of these two regimes can be understood considering the
interplay of vortex and quasiparticle response: at low temperature
flux motion contribution is dominant.  Near the transition temperature
$T_c$ quasiparticles play a significant role in the electromagnetic
response (see Fig.  \ref{sd-like}).  Very close and above $T_c$
thermal fluctuations set in.  In absence of disorder, all those
mechanisms merge smoothly one into the other: in fact, the low
temperature limit of the fluctuations-dominated dc resistivity
coincides with the free flux flow \cite{IkedaJPSJ91}, and a nearly
frequency-independent $\rho_{1}$ results for both the fluctuation
induced dissipation
\cite{MikeskaZP70,SchmidtZP,SkocpolRPP75,KlemmJLTP74,DorseyPRB91,
SilvaEPJB02} and for flux (free) flow resistivity (as can be easily
seen from Eqs.(\ref{CCformulae}), considering that the free flow limit
corresponds to the case $\epsilon=1$).

\begin{figure}[t]
\begin{center}
\begin{minipage}[h]{80mm}
\epsfig{file=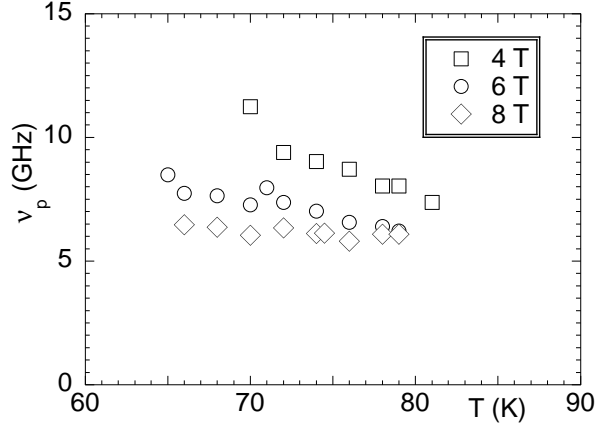, width=80mm}
\end{minipage}
\caption{Temperature dependence at various fields of the depinning
frequency $\nu_{p}$ in the temperature region below $T_{f}$. It is 
seen that $\nu_{p}<$ 15 GHz, at all fields and temperatures.}
\label{YCorb-nup}
\end{center}
\end{figure}

This framework drastically changes in presence of disorder.  In that
case vortices are more or less pinned to defects, and the overall dc
resistivity becomes smaller than the free flow expression
(viscous/plastic regime), eventually reaching a zero value if the
interactions are strong enough to completely lock the flux lines (frozen
regime).  For what concerns the frequency response, the
viscous/plastic as well as the frozen regime are characterized by a
value of the creep factor $\epsilon$ (see Eqs.(\ref{CCformulae}))
lower than unity (zero in the frozen regime), thus resulting in a
frequency dependent real resistivity.  In particular,
$\mathrm{Re}[\rho_v]$ should grow from $\rho_{dc}$ to the free flow
value $\rho_{ff}$ when the frequency increases across the
depinning frequency $\nu_p$.  This pinning frequency is estimated to be
$\nu_p \simeq 10$ GHz \cite{WuPRL95,TsuchiyaPRB01,SartiPhC04} in YBCO,
well within the frequency range here explored.  As a result, the large
frequency dependence of our data below $T_f$ is explained with the
increased effect of disorder.
\begin{figure}[t]
\begin{center}
\begin{minipage}[h]{80mm}
\epsfig{file=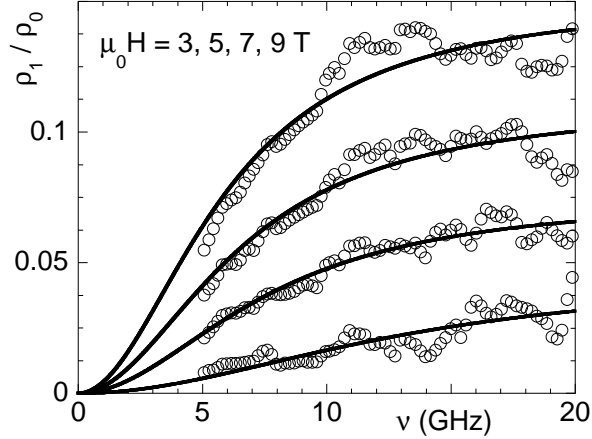, width=80mm}
\end{minipage}
\caption{Typical swept-frequency measurements at different fields and
$T=$ 70 K. A marked frequency dependence is observed, denoting the
change of vortex response.  Continuous curves are fits with
Eq.(\ref{CCformulae}).}
\label{YCorb-r-nu-lowT}
\end{center}
\end{figure}
We now turn to the analysis of the data deep in the strongly
frequency-dependent regime, that is $T<T_{f}$.  We report in
Fig.\ref{YCorb-r-nu-lowT} typical frequency sweeps for the normalized
real resistivity in sample Y3 at low temperature and at various
magnetic fields.  It is immediately apparent that
$\mathrm{Re}[\rho_{1}]$ increases as a function of frequency,
eventually reaching a saturation value at high frequencies.  This is
exactly the behavior expected from mean-field theories of the vortex
state at sufficiently low temperatures.  Within this framework, the
field variation of the real resistivity is entirely given by the
vortex motion, so that we fitted the data with the first of
Eq.(\ref{CCformulae}), using $\epsilon$, $\nu_0$ and
$\rho_{ff}/\rho_{n}=\Phi_{0}B/\eta\rho_{n}$ as fitting parameters.  As
it can be seen (thick lines), good fits are obtained.  Similar results
are obtained at different temperatures and fields.  It should be
mentioned that at low fields, and/or at high enough temperatures, the
experimental curves become nearly featureless, and fitting is less
reliable.  We report here the vortex parameters obtained in the most
reliable temperature and field ranges, that correspond to the region
close to the frozen regime ($\epsilon\simeq$ 0, $\nu_0\simeq \nu_p$).
It is found that $\nu_p$ smoothly depends on temperature and magnetic
field, as reported in Fig.(\ref{YCorb-nup}), and decreases for both
increasing field and temperature.

\begin{figure}[t]
\begin{center}
\begin{minipage}[h]{80mm}
\epsfig{file=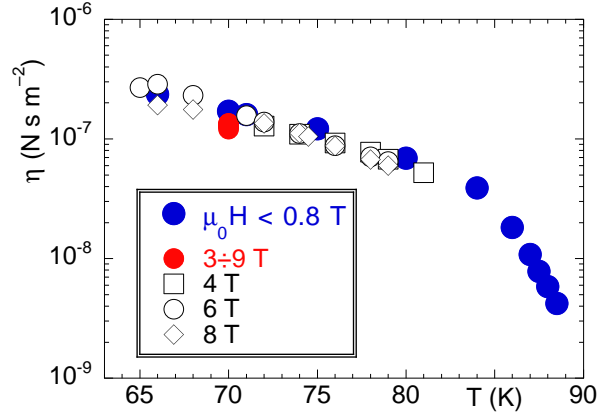, width=80mm}
\end{minipage}
\caption{Temperature dependence of the vortex viscosity $\eta$ in
YBCO, obtained from Corbino disk swept frequency method
(open symbols and red full dots, fields depicted in the figure) and
from field-sweeps, resonant cavity technique (blue full dots).  All
the sets of data coincide, indicating that the intrinsic viscosity is
measured.  Moreover, no significant magnetic field dependence is
detected.}
\label{Yeta}
\end{center}
\end{figure}

The vortex viscosity, as obtained from swept frequency measurements
in the low temperature region is reported in Fig.(\ref{Yeta}). It is
remarkable that no significant magnetic field dependence is detected,
thus indicating that the flux flow follows a magnetic-field linear
dependence, $\rho_{ff}\sim B$.

\begin{figure}[t]
\begin{center}
\begin{minipage}[h]{80mm}
\epsfig{file=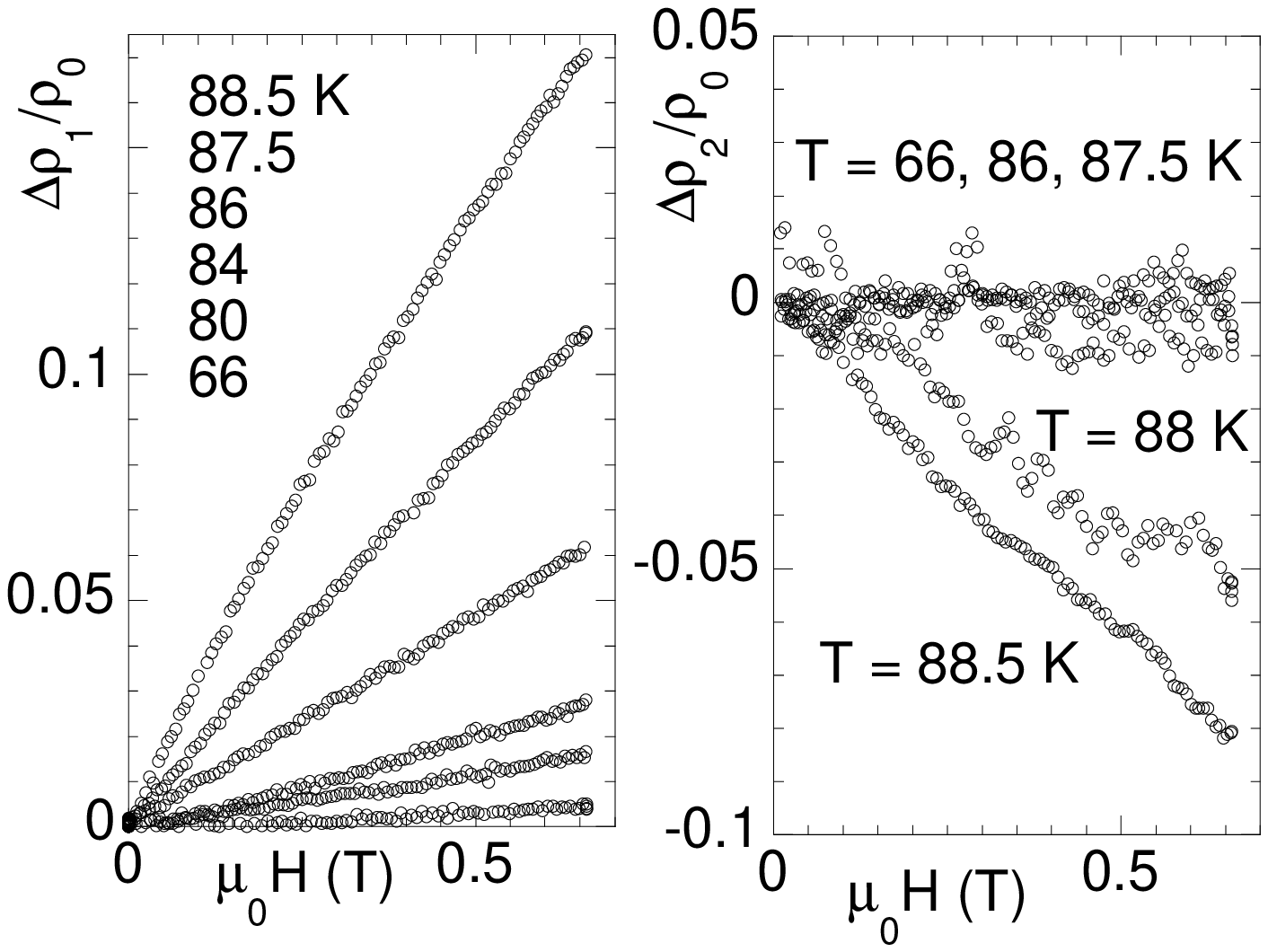, width=80mm}
\end{minipage}
\caption{Typical magnetic field dependence of the variation of the
complex resistivity in YBCO, measured at 48 GHz.  Left panel, real
resitivity. Right panel, imaginary resistivity.  Note at high
temperature the downward curvature in $\Delta\rho_{1}(H)$, and the
decrease of $\Delta\rho_{2}(H)$.}
\label{YCavSweeps}
\end{center}
\end{figure}

To gain more insight in the electromagnetic response at microwave
frequencies, we present now measurements of the complex resistivity
obtained with the cavity resonator at a much higher frequency,
$\nu=\omega/2\pi=$ 48 GHz, as a function of the applied magnetic field on
YBCO samples from the same batch.  Measurements span the temperature
range 65 K $-T_{c}$, and the field is limited to 0.8 T. In
Fig.(\ref{YCavSweeps}) we show typical field-sweeps of the variation of
the complex resistivity at some significant temperature.  It is seen
that the real resistivity $\rho_{1}$ has an almost linear variation
with the applied field, apart possibly the low field region, and that
this linear dependence changes to a concave downward behavior
approaching closely $T_{c}$.  At the same time, the variation of the
imaginary part $\Delta\rho_{2}$ with the field is nearly absent at low
temperatures, and is {\it negative} at higher temperatures, close to
$T_{c}$.

This behavior is easily recognized as an essentially free flux flow,
with a possibly significant contribution of the field-induced
superfluid depletion at high temperatures.  This is completely
consistent with the results from Corbino disk measurements, if one
considers that $\nu_0$ was found (at low temperature) below 15 GHz.
As a consequence $\left(\nu/\nu_{0}\right)^{2}\gg$1 and,
following Eq.(\ref{CCformulae}), the vortex contribution reduces, as a
first approximation, to the real flux flow term.  Then, not too close
to $T_{c}$, where the field dependence is essentially linear,
$\rho_{1}\simeq\rho_{ff}=\Phi_{0}B/\eta$ and $\rho_{2}\simeq$ 0, from
which the vortex viscosity is readily obtained.  Such data points are
reported in Fig.(\ref{Yeta}) and compared to the Corbino disk results.
The agreement is excellent: field sweeps and frequency sweeps give
exactly the same vortex viscosity.

\begin{figure}[t]
\begin{center}
\begin{minipage}[h]{80mm}
\epsfig{file=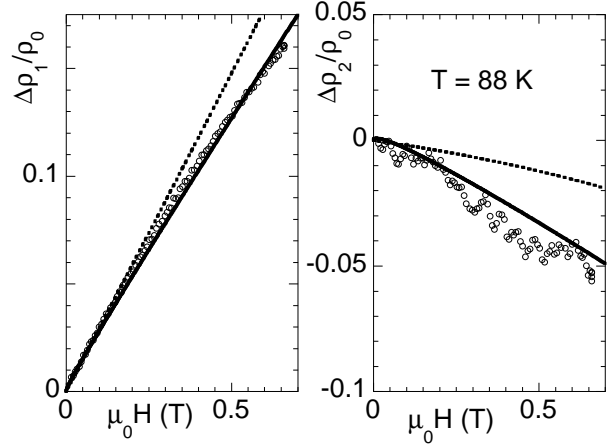, width=80mm}
\end{minipage}
\caption{Typical fits of the magnetic field dependence of the
variation of the complex resistivity in YBCO, measured at 48 GHz, with
the flow expression, Eq.(\ref{genform_hf}).  Continuous lines:
superconductor with lines of nodes, dashed lines: fully gapped
superconductor.}
\label{YCavFits}
\end{center}
\end{figure}

At higher temperatures both the downward curvature of
$\Delta\rho_{1}(H)$ and the pronounced, negative $\Delta\rho_{2}(H)$
can be accounted for by the field dependence of the superfluid and
quasiparticle concentration.  As discussed in Sec.\ref{theory}, in
this case one has to use the full expression of the vortex state
resistivity, Eq.(\ref{genform}), with the noticeable simplification
that $\tilde\rho_{v}\simeq\rho_{ff}$ due to the high operating
frequency.  Fitting necessarily add other parameters: the zero field,
temperature-dependent superfluid fraction $x_{s0}(T)$, the
zero-temperature London penetration depth $\lambda_{0}$, the exponent
$\alpha$ and the factor $c$ (See Eq.(\ref{nodi})).  The upper critical
field is not a different fitting parameter, assuming
$\eta=\Phi_{0}B_{c2}/\rho_{n}$.  We were not able to fit the pairs of
curves $\Delta\rho_{1}(B)$, $\Delta\rho_{2}(B)$ with the exponent
$\alpha$=1 typical of fully-gapped wavefunction.  Instead, by using
$\alpha=\frac{1}{2}$ typical of a wavefunction with lines of nodes in
the gap \cite{MallozziPRL98,VolovikJETP93,WonPRB96}, we obtained fits
as reported in Fig.(\ref{YCavFits}) with the reasonable choice
$\lambda_{0}$= 160 nm, and $c\simeq$ 0.15 (we found $c\simeq$ 1 in a
different sample, with the analysis of the real part only
\cite{SilvaSUST00}).  The resulting $\eta$ connects smoothly to the
experimental data obtained at lower temperatures, and
$x_{s0}(T)\sim\left[1-\left(\frac{T}{T_{c}}\right)^2\right]$.  While
this is not a direct evidence for unconventional superconducting gap,
we note that by taking into account the possible existence of lines of
nodes (and the consequent increased weakness of the superfluid
stiffness), the vortex state microwave response is fully described in
its temperature, frequency and field dependence from well below up to
very close to $T_{c}$.  It is interesting to notice that the small
quasiparticle contribution observed (apart very close to $T_c$) is
consistent with $\tau_{qp}\sim\tau_n$
(see Fig.\ref{sd-like}).  Our results suggests then that in
our YBCO films $\tau_{qp}/\tau_n\simeq 1$, closer to the results of 
\cite{GaoPRB96} rather than to those of \cite{NagashimaFujita94}.
In this case YBCO is a paradigmatic
case, and the simplest to understand.  We will see in the following
that all the other materials below reported present additional
physics, that prevent from a straightforward application of the CC
model.  Additional mechanisms are revealed by measurements of the
microwave response.
\subsection{MgB$_{2}$}
\begin{figure}[t]
\begin{center}
\begin{minipage}[h]{80mm}
\epsfig{file=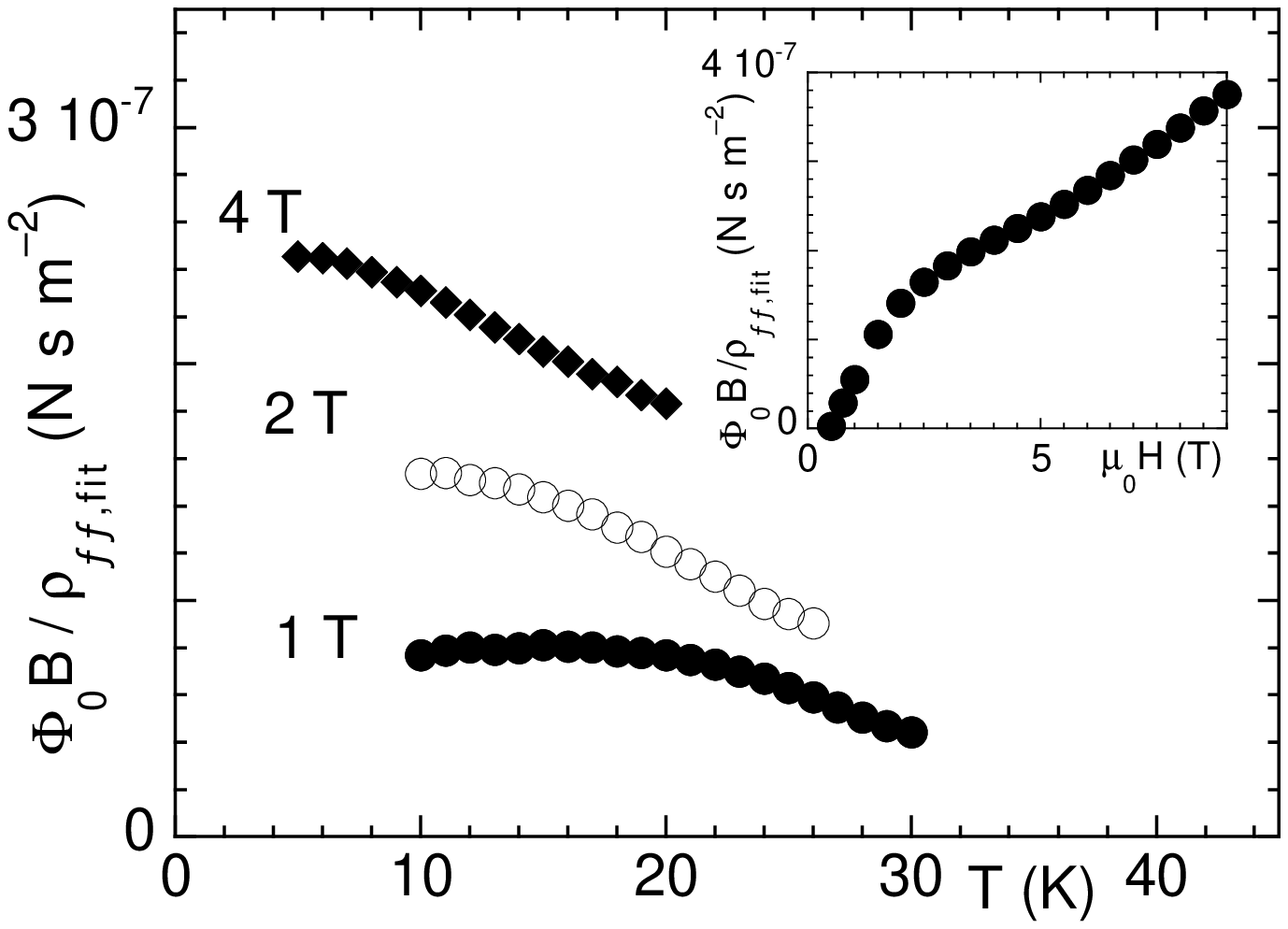, width=80mm}
\end{minipage}
\caption{Field and temperature dependence of the ratio
$\Phi_0B/\rho_{ff,fit}$ as obtained from the MgB$_2$ data fitted through
Eq.(\ref{CCformulae}).  This ratio should equals $\eta$. It is readily
seen that both temperature and
magnetic field dependences present very exotic behavior.}
\label{MCorb-wrongparam}
\end{center}
\end{figure}

\begin{figure}[t]
\begin{center}
\begin{minipage}[h]{80mm}
\epsfig{file=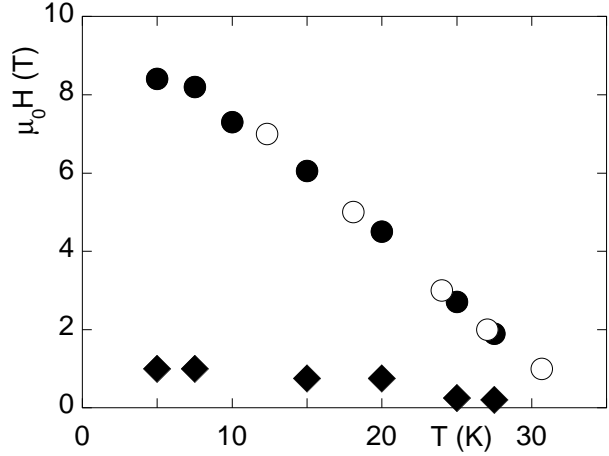, width=80mm}
\end{minipage}
\caption{$H-T$ phase diagram in MgB$_{2}$. Open circles: $H_{c2}$ from 
\textit{dc} measurements. Full dots, $H_{c2}$ from the selfconsistent 
fitting procedure. Diamonds, $H_{1}$.}
\label{MCorb-HT}
\end{center}
\end{figure}

It is somehow expected that MgB$_{2}$, due to its firmly established
two-gap nature, can hardly be described by oversimplified models.
\begin{figure}[t]
\begin{center}
\begin{minipage}[h]{80mm}
\epsfig{file=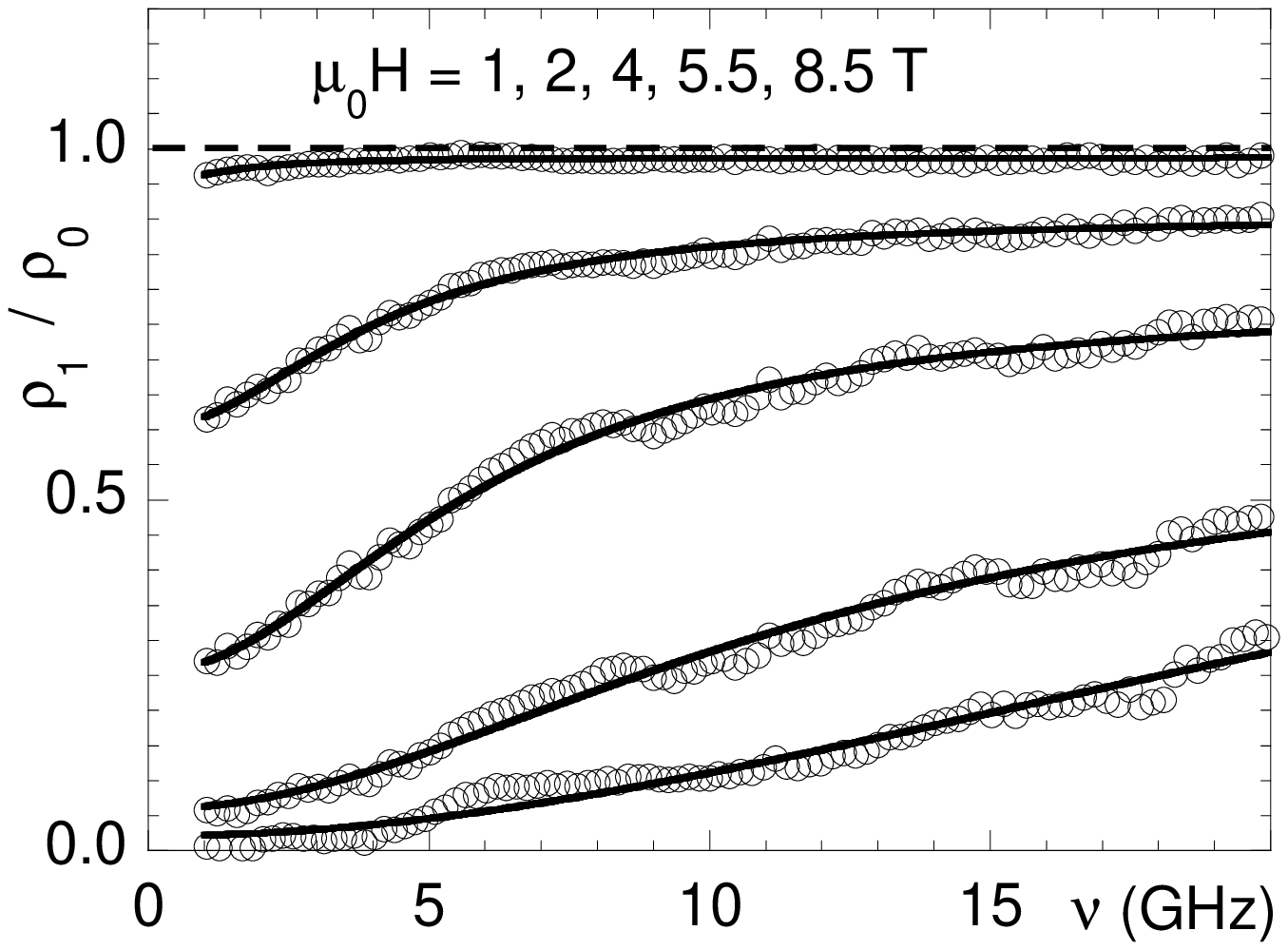, width=80mm}
\end{minipage}
\caption{Frequency dependence of $\rho_{1}$ in MgB$_2$, at $T=$ 10 K
and different magnetic fields (analogous to Fig.\ref{YCorb-r-nu}).
Remarkably, a blind fit with Eq.(\ref{CCformulae}) gives good agreement
(thick continuous lines). However the resulting parameters exhibit
very contradictory field and temperature dependences, see text and
Fig.\ref{MCorb-wrongparam}.}
\label{MCorb-r-nu}
\end{center}
\end{figure}
In order to didactically describe the difficulties hidden in the
analysis of the data in this two-gap compound, we first present in
Fig.(\ref{MCorb-r-nu}) the real resistivity as a function of the
frequency at fixed temperature and various magnetic fields, in
complete analogy with the data for YBCO reported in
Fig.(\ref{YCorb-r-nu}).  The reported data are qualitatively similar
to those measured on YBCO, showing at each field a (real) resistivity
increasing as a function of frequency, reaching a plateau value
$\rho_{pl}$ at high frequencies.  Moreover, it is shown that, rather
surprisingly, the simple, single-gap CC model quantitatively fit the
data.  One might then be led to the conclusion that simple
vortex motion, as given by Eq.(\ref{CCformulae}), captures the physics
involved in microwave response in MgB$_{2}$.  However, despite the quality of
the fits, the behavior of the parameters is not easily understood
within simple models for vortex motion: as an example, we report in
Fig.(\ref{MCorb-wrongparam}) the ratio $\Phi_0 B/\rho_{ff,fit}$ that
should coincide, in this oversimplified model, to the vortex viscosity
$\eta$.  It is found that $\Phi_0 B/\rho_{ff,fit}$ depends very
strongly on the applied field (as opposed to the behavior
theoretically expected and experimentally found in YBCO), while its
temperature dependence changes with the field, becoming less and less
temperature dependent as the field increases.
\begin{figure}[t]
\begin{center}
\begin{minipage}[h]{80mm}
\epsfig{file=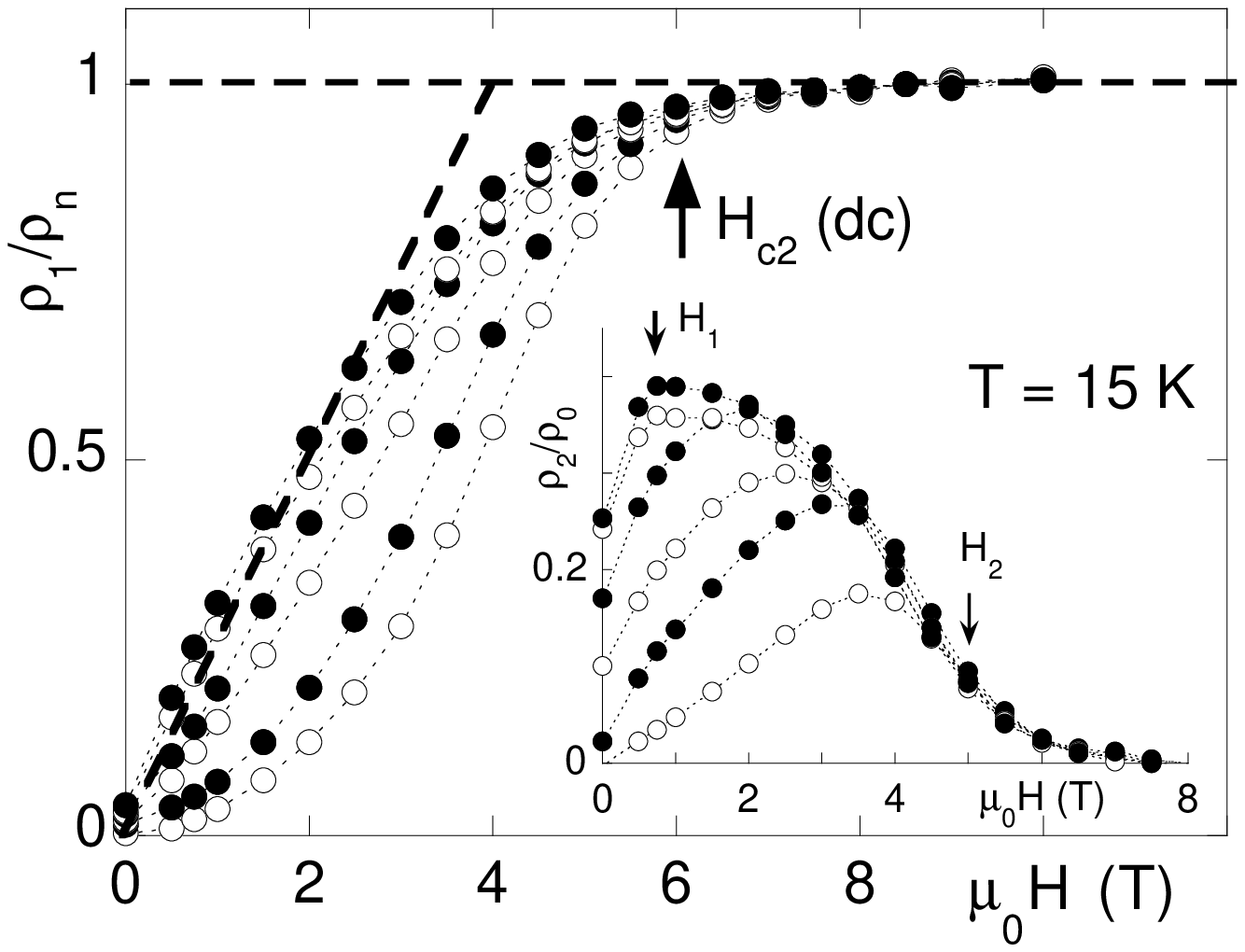, width=80mm}
\end{minipage}
\caption{Field dependence of the complex resistivity at fixed
frequencies $\nu=$2, 5, 9, 12, 15, 18 GHz (frequency increases from
the bottom curve to the top one) and $T=$ 15 K in MgB$_{2}$.  Main
panel: $\rho_{1}/\rho_{0}$.  The intersection of the dashed lines
indicates the upper critical field as it would result from a naive
application of a vortex motion model.  Inset: $\rho_{2}/\rho_{0}$.
Arrows indicate the field $H_{1}$ where the high-frequency imaginary
resistivity reaches a maximum and the field $H_{2}$ where there is no
more a clear frequency dependence in $\rho_{2}$.}
\label{MCorb-r-H}
\end{center}
\end{figure}
More puzzling results can be revealed by reporting the normalized
complex resistivity as a function of magnetic field, at fixed
temperature and different frequencies.  In Fig.(\ref{MCorb-r-H}) we
report these data at $T=15$ K, but similar results are obtained at all
temperatures \cite{SartiPhC04}.  From this figure, it is immediately
apparent that the straightforward application of a single-gap model,
such as the CC model in its original formulation, leads to
considerable contradictions.  First of all, should one identify the
initial, $B$-linear part of $\rho_{1}$ at the highest frequency with
the flux flow resistivity $\rho_{ff}$, in analogy with the
field-sweeps in YBCO, the resulting $H_{c2}$ as obtained from the
linear extrapolation depicted in Fig.(\ref{MCorb-r-H}) would be a
factor 1.5-2 lower than $H_{c2}$ independently measured on the same
sample by dc resistivity \cite{FerrandoPRB03}.  Second, at low fields
and high frequency there is a steep increase in $\rho_{2}(H)$, up to a
smoothly temperature-dependent field $H_{1}$, as reported in
Fig.\ref{MCorb-HT}.  Taking the vortex motion as the only source
of response, this is possible only with the assumption that
$\nu_{0}\approx$ 15 GHz, and strongly field-dependent.  However, if
this were true, one should observe a large variation (as a function of
frequency) of $\rho_{1}(H)$ in the same field and frequency ranges.
There is no indication of such a strong difference
between the experimental curves $\rho_{1}(H)$ measured at different
frequencies for $\nu>10$ GHz, indicating that typical vortex
frequencies are not placed in this range.  The steep increase of the
imaginary response, followed by the more gradual decrease, is then
more likely ascribed to an unconventional increase of the screening
length, as it could be given by the fast suppression of the superfluid
fraction in the weak $\pi$ band \cite{SartiPRB05}.  We finally note
that, above a crossover field $H_2$, $\rho_2$ is nearly independent on
frequency: there are at present no indications on the possible origins for
this feature, so that in the following we confine the discussion to
fields $H<H_2$.

\begin{figure}[t]
\begin{center}
\begin{minipage}[h]{80mm}
\epsfig{file=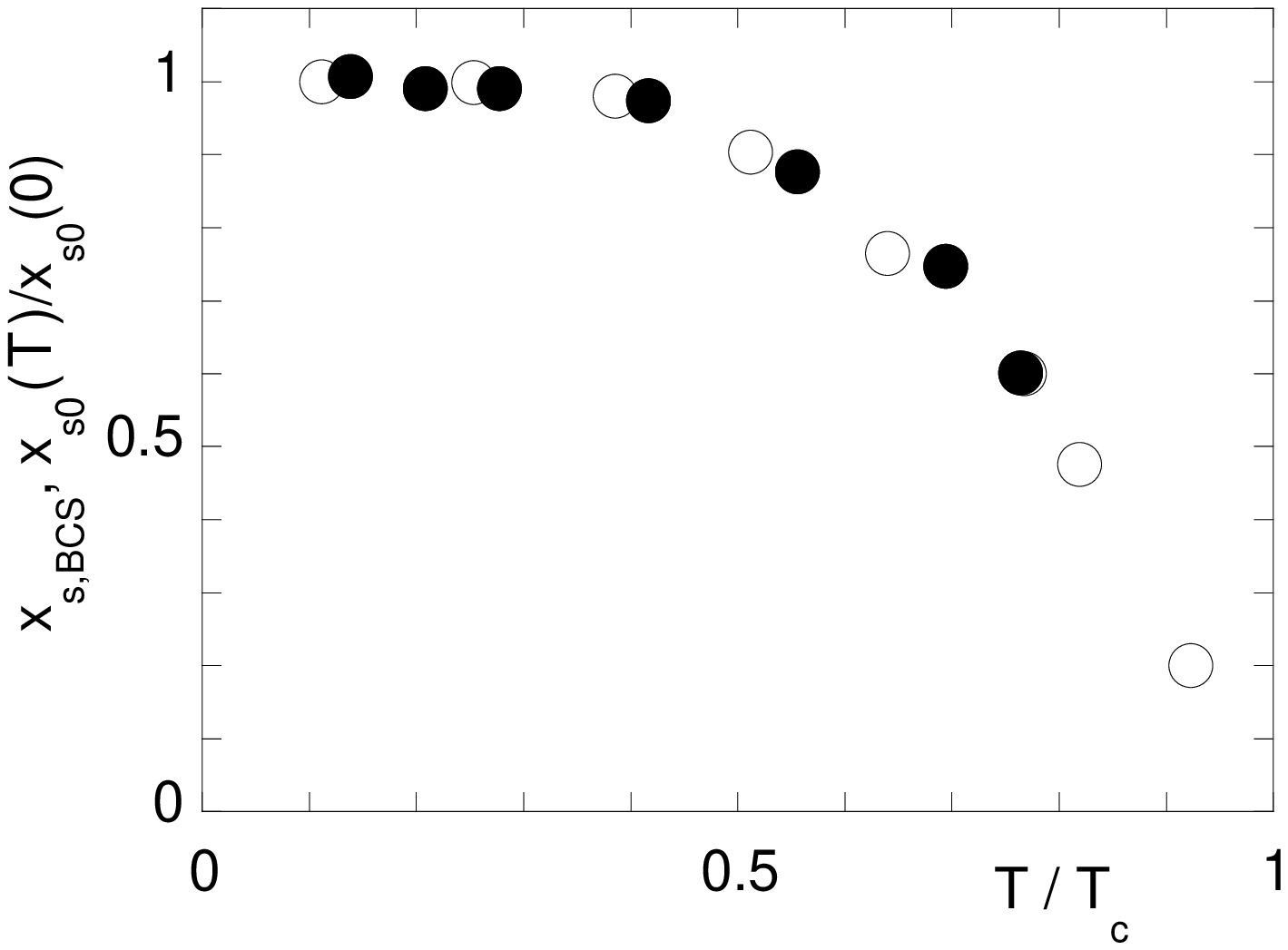, width=80mm}
\end{minipage}
\caption{Full dots: temperature dependence of the zero-field $\sigma-$band
superfluid fractional density obtained from the selfconsistent 
fitting procedure. Open symbols are calculations with the
standard BCS relation \cite{TinkhamINTROSUP} using gap values from
\cite{SzaboPRL01}.}
\label{MCorb-ns}
\end{center}
\end{figure}

We can thus conclude that swept-frequency measurements reveal a
clearly richer physics of the vortex-state in MgB$_{2}$ than, e.g., in
YBCO. Accordingly, the interpretation of the data must be based on
more complex models.

In this perspective, it is interesting to compare the measured
behaviors of $\rho_1(H)$ and $\rho_2(H)$ with the predictions of the
model that takes into account both vortices and quasiparticles,
Eqs.(\ref{genform2}) and Fig.\ref{sd-like}, left panel (we assume
that, according to literature, MgB$_2$ is an s-wave superconductor
\cite{SeneorPRB01}).  By comparing the measured data with the
predicted curves, it is clear that the motion of vortices cannot take
into account the whole observed high frequency behavior.  Indeed, a
rather large quasiparticle contribution (curves with small $x_{s0}$)
is needed to obtain behaviors similar to those observed: it is seen
from Fig.\ref{sd-like}, left panel, that (at high frequency: in
Fig.\ref{sd-like} it is assumed $\nu\gg\nu_0$) $\rho_1(H)$ should
increase approximately linearly with field while $\rho_2(H)$ should be
more or less constant at low fields and then decrease approaching the
upper critical field.  Apart from the low field region ($H<H_1$), this
behavior is rather similar to the measured one.  We can then conclude
that, provided a large quasiparticle contribution is taken into
account, the behaviors observed could possibly be described in terms
of rather simple one-band models for $H>H_1$.  This result is fully
consistent with spectroscopic measurements
\cite{EskildsenPRL02,GonnelliPRB04} which show that the superfluid
density of the $\pi$ band decreases strongly with the magnetic field,
nearly vanishing at a characteristic field $H^{*}$: the microwave
response in the vortex state at low fields, where both gaps play a
role, cannot be described by a singe gap model, while the higher field
data can be nicely described by that model.
This picture is reinforced by noting that both the temperature
dependence and the numerical values of $H_1$ agree very well with
those reported for $H^*$
\cite{EskildsenPRL02,GonnelliPRB04,CubittPRL03,KoshelevPRL03}.

We now show that the microwave resistivity above $H_{1}(T)$
quantitatively coincides with a model for a single-band superconductor,
Eqs.(\ref{genform2}), if the conductivity of the $\pi$ band is taken
as purely real.

We make some rather crude approximations in order to reduce the number
of fitting parameters.  First of all, we assume that above $H_1$ the
residual $\pi-$band contribution to the superfluid can be neglected,
so that the superfluid fraction $x_s(T,B)$ is related to the
superfluid fraction of the $\sigma$ band only, $x_{s,\sigma}(T,B) =
N_{s,\sigma}/N_\sigma$ (that is, the superfluid volume density divided
by the total volume density of electrons in the $\sigma$ band),
through the relation $x_s(T,B) = K x_{s,\sigma}(T,B)$, with $K =
N_\sigma/(N_\sigma+N_\pi)$, being $N_\pi$ the volume density of
electrons in the $\pi$ band.  The penetration depth above $H_{1}$
becomes then
$\lambda^2(T,B)=\lambda_{0,\sigma}^2/x_{s,\sigma}(T,B)$.

For what concerns the normal fluid resistivity
$\rho_{nf}=1/\sigma_{nf}$, we assume that, since the $\sigma$ and
$\pi$ bands interact very weakly with each other \cite{MazinPRL02},
above $H_1$ one can write $\sigma_{nf} \simeq
\sigma_{n,\pi}+[1-x_{s,\sigma}(T,B)]\sigma_{n,\sigma}$.  Using the
independently measured \cite{FerrandoSUST03,FerrandoCM03}, temperature
dependent $H_{c2}$ anisotropy of our sample, the Gurevich model
\cite{GurevichPRB03} yields $\sigma_{n,\sigma}/\sigma_{n,\pi} \simeq
0.25$.  One then obtains, to a good approximation,
$\rho_{nf}/\rho_n=1$ in all the field and temperature region
$H>H_1(T)$, and the $T$ and $B$ variations of $\nu_s(T,B)$ are then
entirely determined by $x_s(T,B)=x_{s0}(T)(1 - b)$, where the last
equality comes from the single-gap-like behavior above $H_{1}$,
$b=B/B_{c2}(T)\simeq H/H_{c2}(T)$ and $x_{s0}(T)=
Kx_{s,\sigma}(B=0,T)$ is the (extrapolated) zero field
value\footnote{We remark that $x_{s0}(T)$ is different from the value
that is obtained through experiments that measure the superfluid
density at $B=0$, being in that case $x_s^{meas}(B=0,T) =
x_{s,\sigma}(B=0,T)+x_{s,\pi}(B=0,T)$} of $x_s$.
Once this general frame has been established, it is possible to
extract from the data, by means of a selfconsistent procedure reported
in the Appendix, the temperature dependent $\sigma$-band superfluid
fraction $x_{s0}(T)$, the temperature dependent upper critical field
$H_{c2}(T)$, the vortex motion resistivity $\rho_{v}(\nu)$ and,
consequently, the characteristic frequency $\nu_{0}$.  We now discuss
those quantities.
In Fig.\ref{MCorb-ns} we report the behavior of $x_{s0}(T)$,
normalized to the lowest temperature value $x_{s0}(T=5$ K$)$.  We
compare the $T$ dependence of $x_{s0}=Kx_{s,\sigma}(B=0,T)$, with
$x_{s,\sigma}(B=0,T)$ as obtained using the expressions of the BCS
theory \cite{TinkhamINTROSUP}, and the values of the gap
$\Delta_\sigma (T)$ measured by point contact spectroscopy
\cite{SzaboPRL01}.  The theoretical calculation and data points from
our microwave measurements agree very well, giving a strong support to
the consistency of the analysis here performed, and adding evidence
that the temperature dependence of the zero-field superfluid fraction
in the $\sigma$ band follows a rather conventional behavior.

Similarly, we obtain values of the upper critical field $H_{c2}$ (Fig.
\ref{MCorb-HT}) in nearly perfect agreement with values obtained
independently from the $dc$ measurements \cite{SartiPRB05}, giving
further confirmation on the reliability and consistency of the
underlying model.  We remark that this is a nontrivial result:
previous analysis have never been able to describe the microwave data
using the same parameters obtained from low frequency measurements.
In particular, the upper critical field was found to be field
dependent \cite{DulcicPRB02} or anomalous field dependencies for the
vortex viscosity were invoked \cite{ShibataPRB03}.

\begin{figure}[t]
\begin{center}
\begin{minipage}[h]{80mm}
\epsfig{file=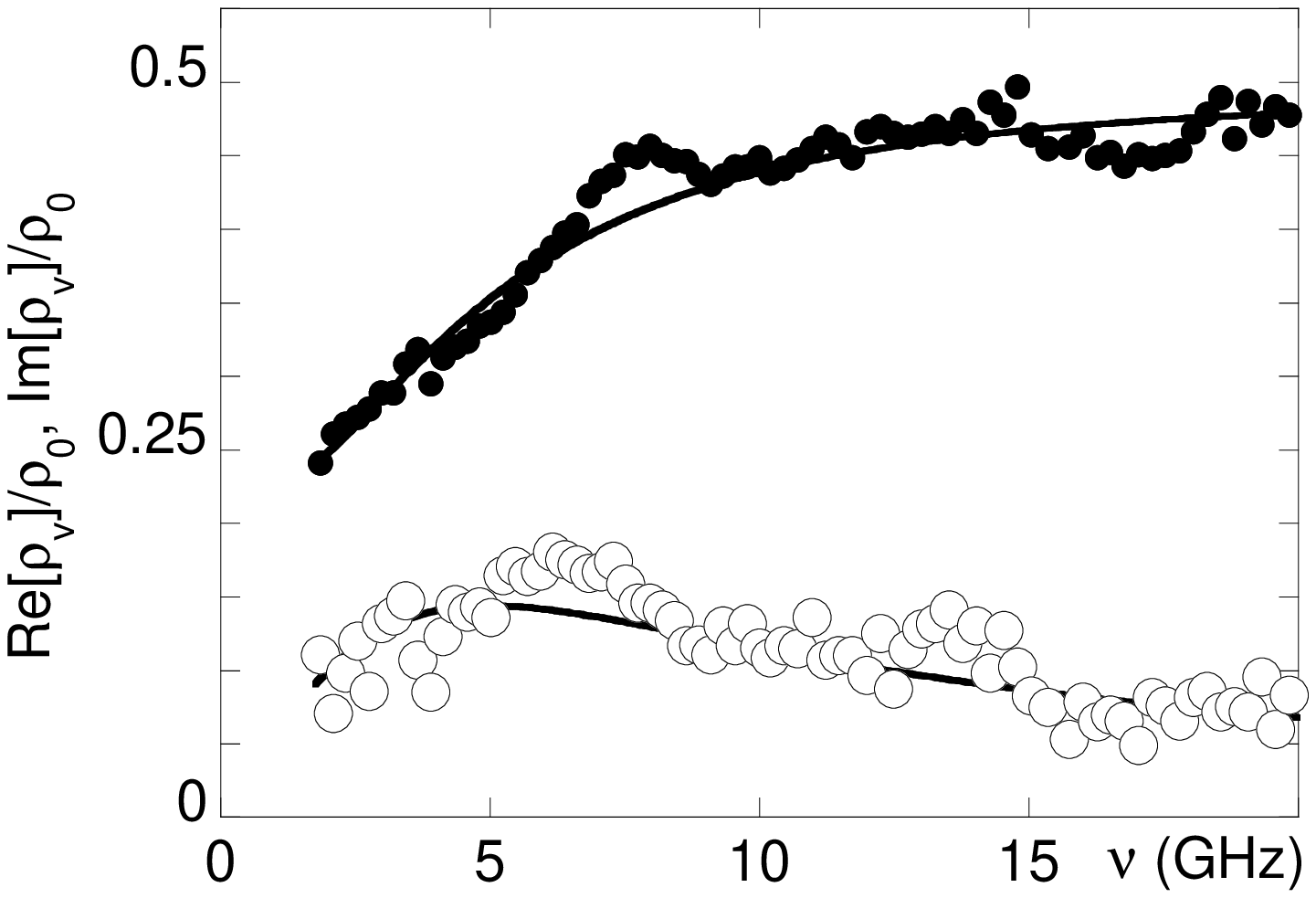, width=80mm}
\end{minipage}
\caption{Contribution of the vortex motion to the overall
field-dependent complex resistivity in MgB$_{2}$, isolated as 
described in the text. Full dots,
$\mathrm{Re}[\rho_{v}]$; open circles,
$\mathrm{Im}[\rho_{v}]$. Continuous lines
are simultaneous fits by Eqs.(\ref{CCformulae}).}
\label{MCorb-vortex}
\end{center}
\end{figure}
\begin{figure}[t]
\begin{center}
\begin{minipage}[h]{80mm}
\epsfig{file=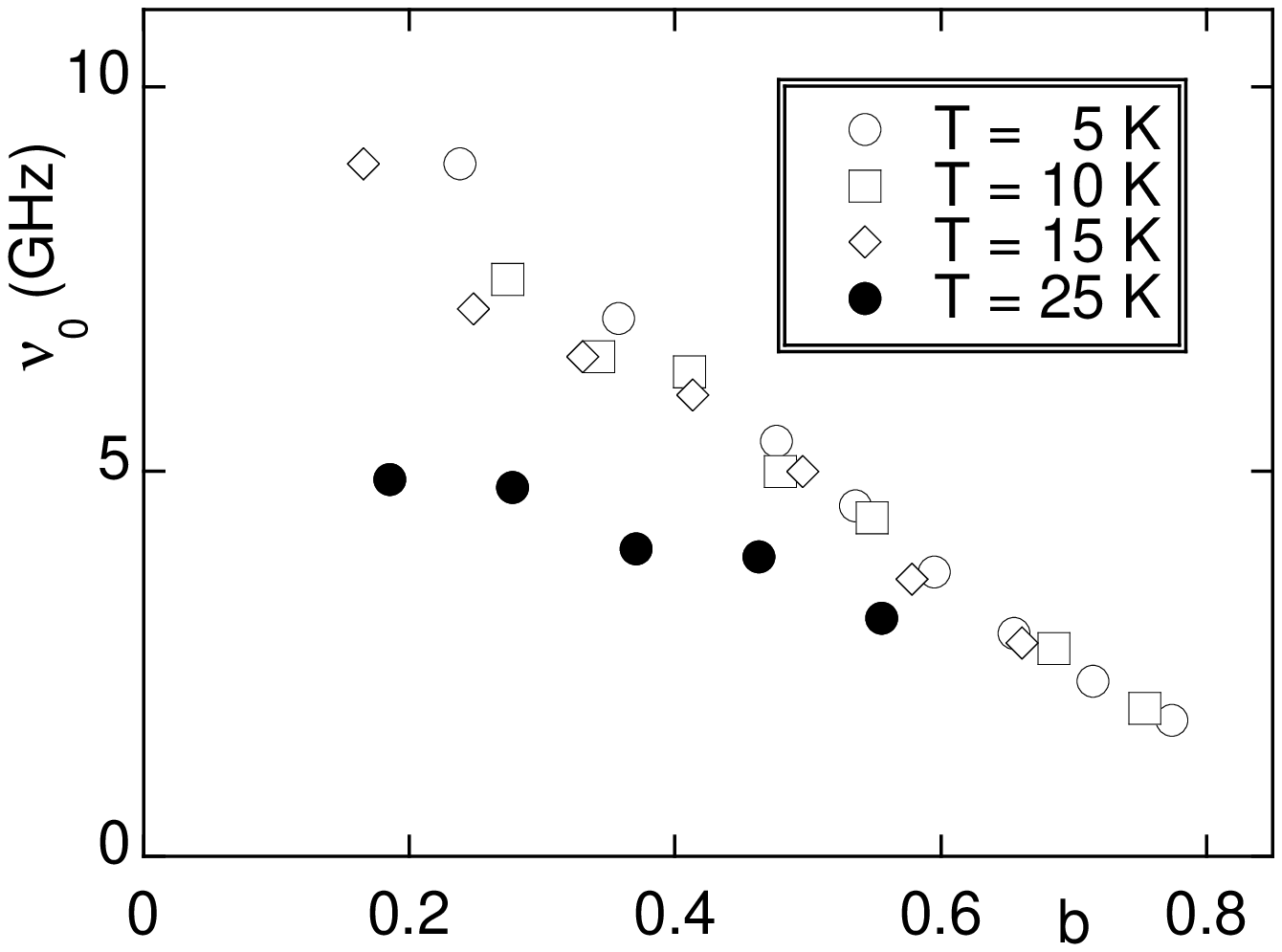, width=80mm}
\end{minipage}
\caption{Reduced field dependence of the vortex characteristic
frequency $\nu_{0}$ in MgB$_{2}$ at various temperatures. The 
reduced field $b=B/B_{c2}$.}
\label{MCorb-nu0}
\end{center}
\end{figure}

Finally we report in Fig.\ref{MCorb-vortex} the vortex motion
complex resistivity, isolated from the experimental data by means of
the selfconsistent procedure.  We remark that now those data are not
flawed, as the raw data, by the presence of the strong contribution of
the $\pi$ band, or of the overall field dependence of the superfluid
concentration (see the Appendix).  We then fitted all the pairs of
curves $\mathrm{Re}[\tilde\rho_v]/\rho_n (\nu)$,
$\mathrm{Im}[\tilde\rho_v]/\rho_n (\nu)$ to Eqs.(\ref{CCformulae}) at
all given temperatures and fields, using $\epsilon$ and $\nu_0$ as
fitting parameters ($\rho_{ff}/\rho_n = H/H_{c2}$ is determined by the
value of $H_{c2}$ obtained previously).  A typical fit for $T = 15$ K
and $\mu_0H=3$ T is reported in Fig.(\ref{MCorb-vortex}).  We find,
consistently with the indication given by the small frequency
dependence of $\rho_{1}$, that $\nu_0$ never exceeds 10 GHz, for any
temperature and field.  In addition, we find a rather strong field
dependence of $\nu_0$ at different temperatures, as reported in
Fig.(\ref{MCorb-nu0}).  This is an indication, in agreement with
previous findings \cite{DulcicPRB02}, of the collective nature of the
pinning forces in this material.  This conclusion is also supported by
the low temperatures collapsing of all curves $\nu_0(b)$, with 
$b=B/B_{c2}(T)$.

Summarizing, we have shown that multifrequency measurements are a key
factor to elucidate the role of quasiparticle, superfluid and vortex
motion in the overall transport properties of MgB$_{2}$ in the vortex
state.  By suppressing the $\pi$ band contribution with a sufficiently
strong magnetic field it has been possible to evaluate the $\sigma$
band superfluid density, the upper critical field and the characteristic
vortex frequency.
\subsection{SmBa$_{2}$Cu$_{3}$O$_{7-\delta}$}
\begin{figure}[t]
\begin{center}
\begin{minipage}[h]{80mm}
\epsfig{file=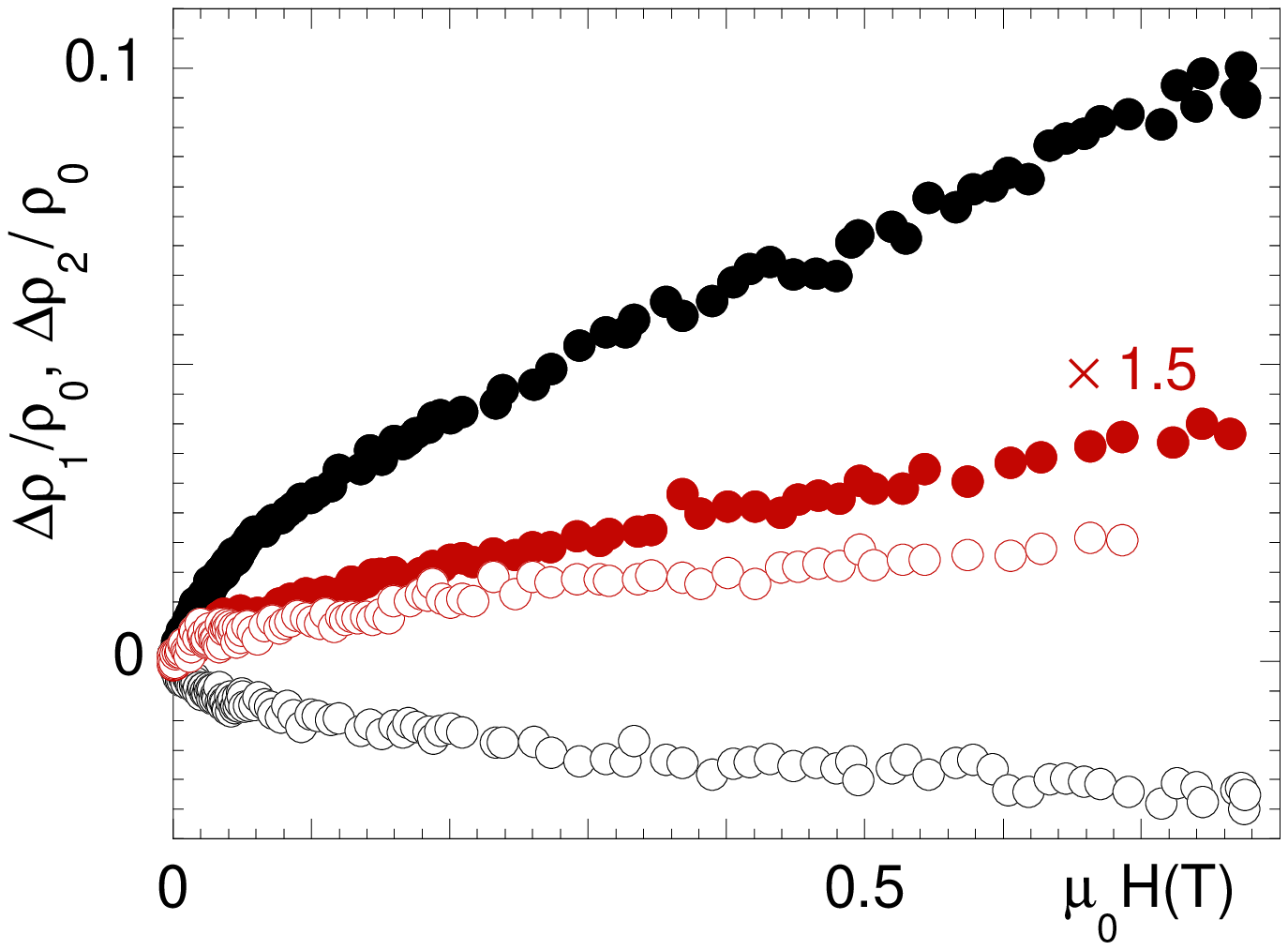, width=80mm}
\end{minipage}
\caption{Field dependence of the complex resistivity changes in SmBCO
(sample Sm2) at two temperatures.  Black symbols, $T=$ 82 K. Red
symbols, $T=$ 71 K. A sublinear component is evident both in
$\Delta\rho_{1}/\rho_{0}$ (full dots) and in $\Delta\rho_{2}/\rho_{0}$
(open circles).  Note the large values and the sign change of
$\Delta\rho_{2}/\rho_{0}$ with increasing temperature. 
$\Delta\rho_{1}/\rho_{0}$ at 71 K has been scaled to avoid crowding.}
\label{SmCav-dat}
\end{center}
\end{figure}

The case of SmBa$_{2}$Cu$_{3}$O$_{7-\delta}$ (SmBCO) is somehow
similar to the case of MgB$_{2}$, in that an additional, strong
contribution different from vortex motion affects the complex
resistivity.  In this Subsection we report measurements of the complex
resistivity for temperatures down to 65 K in moderate fields
$\mu_{0}H<$0.8 T and at the high operating frequency $\omega/2\pi=$ 48
GHz.  Most of the measurements are thus taken below the
irreversibility line of SmBCO \cite{MurakamiSUST96,KupferPRB99}.
Typical field sweeps for the variation of the complex resistivity at
various sample temperatures are reported in Fig.(\ref{SmCav-dat}). The
data here reported are representative of the behavior observed in
other SmBCO films measured under the same conditions and in similar
temperature and field ranges.  We list the most relevant experimental
features that will determine the discussion of the data, and we
specifically compare them to the behavior of the parent compound
YBCO.

\begin{figure}[t]
\begin{center}
\begin{minipage}[h]{80mm}
\epsfig{file=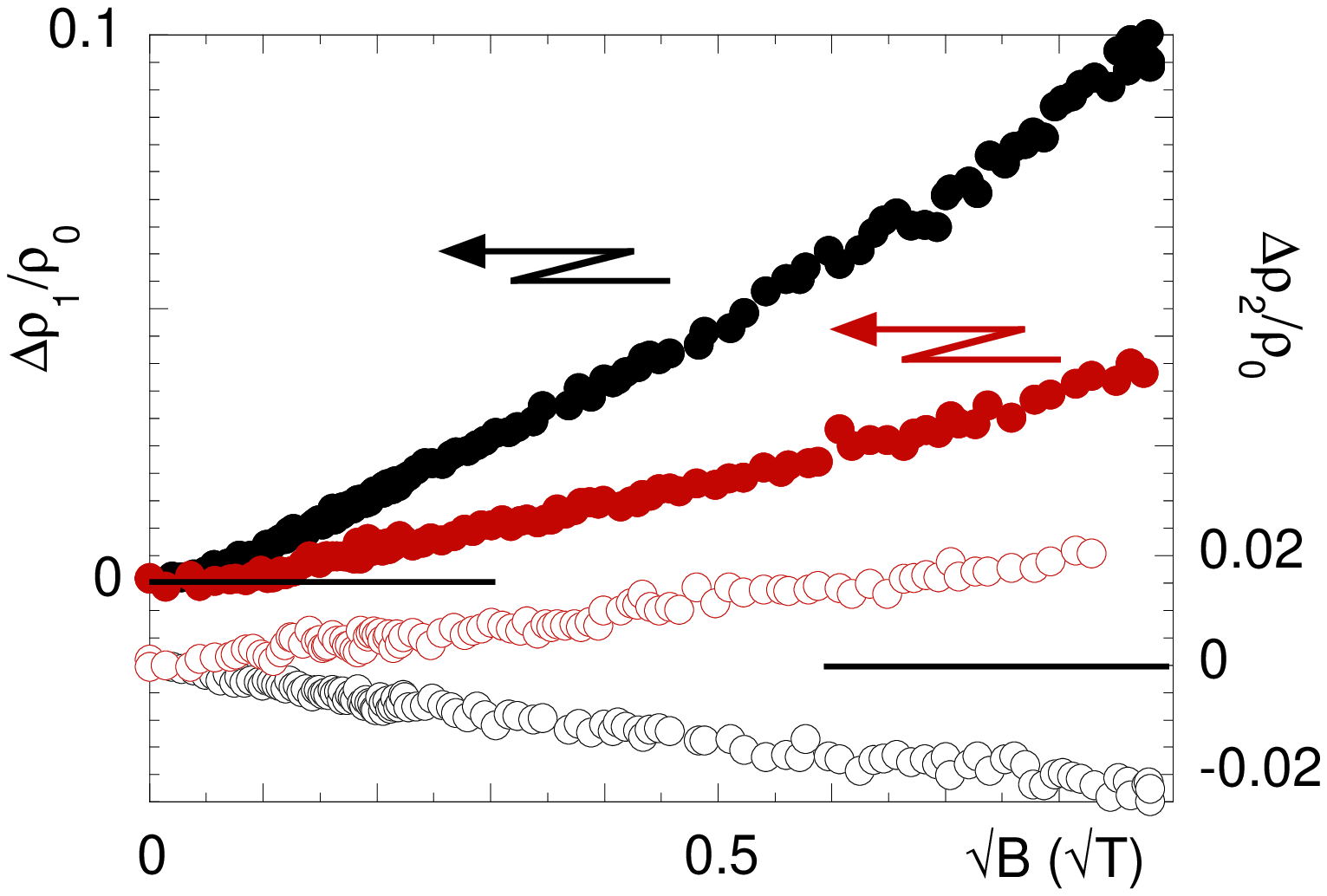, width=80mm}
\end{minipage}
\caption{Same data of Fig.(\ref{SmCav-dat}), replotted \textit{vs.}
$\sqrt{H}$.  Left scale, $\Delta\rho_{1}/\rho_{0}$.  Right scale,
$\Delta\rho_{2}/\rho_{0}$.  Symbols as in Fig.(\ref{SmCav-dat}). It clearly 
appears that $\Delta\rho_{2}/\rho_{0}\propto\sqrt{H}$.}
\label{SmCav-datsqrt}
\end{center}
\end{figure}

At all temperatures, $\Delta\rho_{1}/\rho_{0}$ exhibits a pronounced downward
curvature at low fields, followed by an approximately linear increase
at higher fields.  By contrast, $\Delta\rho_{2}/\rho_{0}$ never shows a linear
variation.  Moreover, $\Delta\rho_{2}/\rho_{0}$ changes from positive
to negative
as the temperature increases, but without changing the curvature of
the data.  Those considerations can be put on more quantitative
grounds by plotting the data as a function of $\sqrt{H}$, as reported
in Fig.\ref{SmCav-datsqrt}. It is seen
that $\Delta\rho_{2}/\rho_{0}$ is well approximated by a straight line, which
corresponds to a $\sim \sqrt{H}$ dependence, while upward curvature in
$\Delta\rho_{1}/\rho_{n}$ \textit{vs.} $\sqrt{H}$ indicates the
presence of both a square
root and a linear term in the $H$ dependence. Summarizing, to the best
of our experimental accuracy, the complex resistivity in SmBCO can be
described by $\Delta (\rho_{1}+\mathrm{i}\rho_{2})/\rho_{0} \simeq
\left[a_{1}(T)+\mathrm{i}a_{2}(T)\right]\sqrt{\mu_{0}H}+b_{1}(T)\mu_{0}H$.

\begin{figure}[t]
\begin{center}
\begin{minipage}[h]{80mm}
\epsfig{file=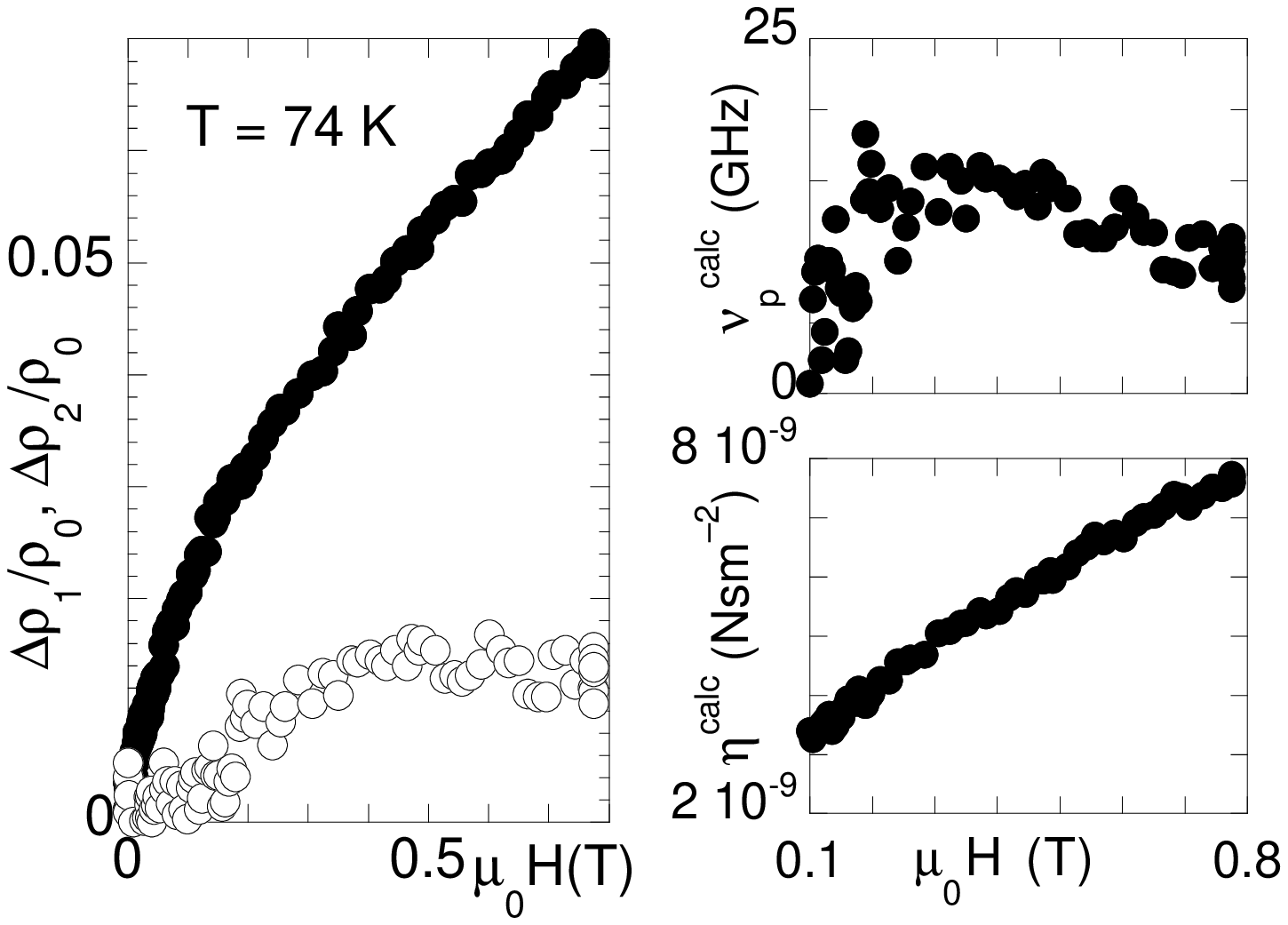, width=80mm}
\end{minipage}
\caption{Left panel: complex resistivity changes at 74 K in sample
Sm1.  Right panels: field dependence of the calculated vortex
parameters $\nu_{p}^{calc}$ and $\eta^{calc}$, as they would result
from the conventional framework for vortex-state complex resistivity.
The calculated viscosity $\eta^{calc}$ presents a strong field
dependence, at odds with models and with experimental data in YBCO.}
\label{SmCav-conventional}
\end{center}
\end{figure}

With respect to YBCO we then encounter two main differences: the field
dependence, which has a strong sublinear component in the entire
temperature range explored, and the very relevant increase of the
imaginary part in even moderate fields.  In order to discuss the data,
in analogy to the discussion on the other materials here investigated,
we first tentatively ascribe the observed behavior to the vortex
motion alone.  For the purpose of the discussion, it is sufficient to
focus on the data at low enough temperature, that is below the
irreversibility line, where creep can be neglected.  We then consider
for the preliminary discussion Eqs.(\ref{gittl-ros}).  In the
conventional data analysis the superfluid density (or, which is the
same, the London penetration depth) does not change appreciably with
the field, and pinning affects mostly the imaginary part of the
response.  One might then be tempted to assign the difference between
YBCO and SmBCO simply to a much stronger, or different, pinning in
SmBCO. Making use of the vortex-motion-only expressions,
Eqs.(\ref{gittl-ros}), one would directly calculate the pinning
frequency from the data\footnote{This kind of analysis has been widely
used in the past in various HTCS, as reviewed in
\cite{GolosovskySUST96}.} as:
$\nu_{p}^{calc}=\nu\frac{\Delta\rho_{2}}{\Delta\rho_{1}}$.  Proceeding
further within the same framework, once the tentative $\nu_{p}$ has
been calculated the viscosity follows immediately by calculating
$\frac{\Phi_{0}B}{\Delta\rho_{1}}\frac{1}{1+\left(\nu_{p}/\nu\right)^{2}}$,
see Eqs.(\ref{gittl-ros}).  However, within this framework we would
obtain a field-dependent vortex viscosity, in particular
$\sim\sqrt{B}$, at odds with the measured vortex viscosity in YBCO.
The procedure is illustrated in Fig.\ref{SmCav-conventional}, where
the as-calculated vortex parameters are reported at one temperature.

Then, it would appear that both the pinning mechanism (implicit in
$\nu_{p}^{calc}$) and the electronic state (implicit in the calculated
vortex viscosity $\eta^{calc}$) are very different in SmBCO and YBCO.
This conclusion does not appear very reasonable, due to the
structural, electrical and superconducting similarities between the
two compounds.  In particular, while pinning may well be
sample-dependent, the much different field behavior of the vortex
viscosity put doubts on the correctness of the simple model used.

We now propose a possible alternative explanation for our data.
Since, as a first approximation, the vortex motion resistivity at low
fields should be proportional to the number of flux lines, that is to
the induction $B\simeq\mu_{0}H$, we tentatively assume that only the
linear part of our data is due to the vortex motion.  In this case, we
deduce from the data that there is almost no vortex motion
contribution to the imaginary resistivity.  This is exactly the
behavior as exhibited by YBCO at the same frequency, that shows free
flux flow at 48 GHz.

\begin{figure}[t]
\begin{center}
\begin{minipage}[h]{80mm}
\epsfig{file=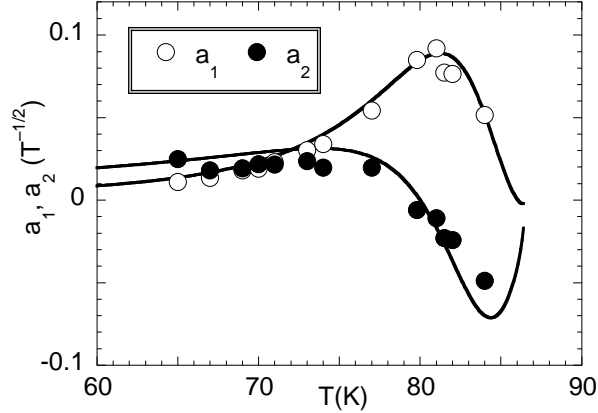, width=80mm}
\end{minipage}
\caption{Plot of the coefficients $a_{1}(T)$ and $a_{1}(T)$.
Continuous lines are simultaneous fits with the pair breaking
expression described in the text.  Details are reported in
\cite{SilvaCM}.}
\label{a1a2}
\end{center}
\end{figure}

We have now to identify the physical mechanism responsible for the
$\sim\sqrt{H}$ part of the real and imaginary resistivity and its
peculiar features (in particular, the change of sign of
$\Delta\rho_{2}(H)$ with increasing temperature).  As shown in Fig.
\ref{sd-like}, right panel, the change of sign of $\Delta\rho_{2}(H)$
with increasing temperature is a signature of the field-dependent
superfluid depletion, while the $\sim\sqrt{H}$ dependence points to a
specific electronic state, namely a superconducting gap with lines of
nodes.  Thus, it is possible to compare theory and experiment.  Making
use of Eq.(\ref{genform2}), we can expand for low fields (small
$H/H_{c2}\simeq B/B_{c2}$) and we obtain explicit expressions for the
coefficients $a_{1}(T)$ and $a_{2}(T)$.  Those expressions contain
several temperature-dependent quantities, for which we believe it is
appropriate to use the simplest possible model: we take the superfluid
fraction $x_{s0}(T)=\left(1-t^{2}\right)$, which is a recognized
approximation in a very wide temperature range for HTCS
\cite{BonnPRB93}, and the pair breaking field $B_{pb}\propto
B_{c2}(T)$, as theoretically suggested \cite{VolovikJETP93,WonPRB96},
so that $B_{pb}=B_{pb0}\left(1-t^{2}\right)$, with $t=T/T_{c}$.  In
order to gain qualitative information, we do not attempt to insert
some temperature dependence in the quasiparticle scattering time
$\tau_{qp}$, that we use as a free parameter.  As can be seen in
Fig.\ref{a1a2}, the fits reproduce the shape, height, width of the
curve given by the experimental data, including the temperature of
change of sign for $a_{2}$.  The resulting $\tau_{qp}\simeq$ 0.7 ps
compares well to the highest values reported in YBCO, e.g., 0.2 ps at
$\approx$ 80 K as obtained from microwave measurements crystals
\cite{BonnPRB93} and to 0.5 ps at $\approx$ 80 K as obtained from
millimeter-wave interferometry in YBCO film \cite{NagashimaFujita94}.
Even if the model can certainly be improved (e.g., by considering a
temperature-dependent $\tau_{qp}$), the substantial agreement with the
data led us to conclude that the microwave resistivity in the
vortex state observed in our SmBCO samples is strongly affected by
the field-induced superfluid depletion, and that the differences
between different samples of YBCO and SmBCO are determined by mere
quantitative differences in $\tau_{qp}$.

Coming back to the fluxon dynamics, once the temperature
dependence of the carrier conductivity has been assessed from the
above described fit, one can extract the vortex viscosity in a
wide temperature range\footnote{At low
temperature the slope $b_{1}$ directly yields $\eta$, but approaching
$T_{c}$ the quasiparticle screening length comes into play and a
correction is needed, see Eq.(\ref{genform}).}. The so-obtained data
points for the vortex viscosity are reported in Fig.\ref{etatutti}.
Both numerical values and temperature dependences compare favorably
to similar data in YBCO (Fig. \ref{Yeta}), adding evidence that the vortex
motion contribution has been correctly identified.

\subsection{General remarks}
\begin{figure}[t]
\begin{center}
\begin{minipage}[h]{80mm}
\epsfig{file=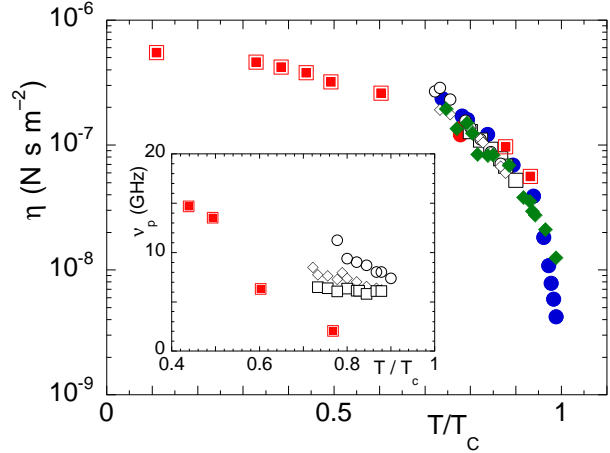, width=80mm}
\end{minipage}
\caption{Temperature dependence of the vortex viscosity $\eta$
obtained in this study in YBCO (data and symbols as in Fig.\ref{Yeta})
and SmBCO films (green filled diamonds), compared to data obtained in
crystals at 19.1 GHz (red squares, from \cite{TsuchiyaPRB01}).  The
data in SmBCO have been multiplied by 2.  All the data collapse on the
same curve, indicating that the same electronic mechanisms take place
in the vortex core.  Inset: depinning frequency in YBCO films (sample
Y3, same data and symbols as in Fig.\ref{YCorb-nup}) compared to
measurements at 19.1 GHz in single crystals (red squares, from
\cite{TsuchiyaPRB01}).}
\label{etatutti}
\end{center}
\end{figure}

As we have shown in this study, the physics of superconductors in the
vortex state can take advantage from investigations at microwave
frequencies. However, in order to assess with some confidence the
various mechanisms responsible for the microwave response, it is
necessary to combine temperature, field and
frequency dependent measurements. This is especially needed in the
two-gap superconductor MgB$_{2}$, where single frequency measurements
are not able to assess the nature of the strong anomaly in the
low-field response.\\
By the combined measurements here reported, it has been shown that, in
all cases here investigated, the correctly identified vortex motion
contribution follows to a great accuracy the conventional models.  In
particular, the high frequency regime always coincides with
conventional flux flow (above $H^{*}$ in MgB$_{2}$): this is a
remarkably noticeable point, since it implies that, in the field and
temperature ranges studied, the dominant excitations inside the
vortex cores in the materials examined are conventional quasiparticle
excitations, as indicated by the temperature dependence and field
independence of the vortex viscosity.  In order to emphasize this
point, we plot in Fig. \ref{etatutti} the vortex viscosity as
measured by us in YBCO and SmBCO thin films and the depinning
frequency obtained by us in YBCO films, together with data obtained in
YBCO crystals by multifrequency cavity measurements
\cite{TsuchiyaPRB01}.  It is
immediately seen that the vortex viscosity has the same behavior in
YBCO and SmBCO, films and crystals.  Moreover, the data for different
materials scale one onto the other with mere numerical factors of order
unity, consistently with the viscosity given by the standard
expression $\eta=\Phi_{0}B_{c2}/\rho_{n}$.  Correctly, the depinning
frequency in YBCO appears to be different in films and in crystals,
indicating that defects in samples (or size effects) play a role.  We
especially emphasize this point, in comparison to earlier studies (see the
early review in \cite{GolosovskySUST96}) that reported the very anomalous
sample independent depinning frequency and sample dependent vortex
viscosity, quite the opposite of the expected behavior.

In fact, a second important indication that emerges from our
measurements is that it is in general not appropriate to neglect the
field dependence of the superfluid density and quasiparticle density
of states in the materials under study: albeit coming from different
physical origins, the superfluid density results to be much weaker
than in conventional superconductors upon the application of an
external magnetic field.  This latter feature seems to be governed by
the (sample-dependent) quasiparticle scattering time in RE-BCO. In
MgB$_{2}$ the low fields behavior is only qualitatively understood,
so that this material appears to be most interesting for future
studies.  In particular, there is at present no consensus on a
representation of the vortex dynamics when both gaps are effective,
which is then a promising field for investigation.
\section{CONCLUSION}
\label{conc}
We have extensively investigated the experimental microwave response
of innovative superconductors (cuprates and MgB$_{2}$) combining
resonant and swept-frequency techniques.  We have shown that this
combination allows for the identification of the major differences
between those superconductors and the conventional, metallic
superconductors.  The main difference resides in the weakness of the
superfluid density with respect to an applied magnetic field both in
cuprates and MgB$_{2}$, coming however from different physics.  By
contrast, the vortex motion contribution appears to be well described
by conventional models (above $H^{*}$ in MgB$_{2}$).
\section*{Appendix A: Selfconsistent fitting procedure in MgB$_{2}$}
Here we describe the selfconsistent fitting procedure for the
swept-frequency data in MgB$_{2}$. We refer to the general 
expressions, Eqs.(\ref{genform}).\\
The values of $x_{s0}$ and $H_{c2}$ for each temperature can be
obtained as follows.  As a first step, we notice that, if
$\frac{\nu}{\nu_s(B,T)} =
2\left(\frac{\lambda(B,T)}{\delta_{nf}(B,T)}\right)^{2}$ is known at
any fixed temperature and field, $r_1(\nu,B,T)$ and $r_2(\nu,B,T)$ can
be obtained from the measured $\rho_1/\rho_n$ and $\rho_2/\rho_n$ by
inverting Eqs.(\ref{genform}).  Although the frequency, temperature
and field dependence of $r_1$ and $r_2$ are not known {\it a priori},
the high frequency limit of $r_1$ is known: at high enough frequency
Eq.(\ref{CCformulae}) gives
$\mathrm{Re}[\rho_v]/\rho_{n}\rightarrow\rho_{ff}/\rho_{n}=b$, with
$b=B/B_{c2}(T)\simeq H/H_{c2}(T)$, having assumed the validity of the
Bardeen Stephen expression \cite{BardeenPR65} for the flux flow
resistivity.  We then choose a given temperature $T_0$ and calculate,
for any value of $H$, a value of $\nu_s(B,T)$ using tentative values
of $x_{s0}(T_0)$ and $H_{c2}(T_0)$ and $x_s(H,T_0)=x_{s0}(T_0)(1-b)$.
We then obtain for all fields $r_1(\nu,B,T_0)$ inverting
Eqs.(\ref{genform}) and we check that they approach a constant value
at high frequency.  The values of $x_{s0}(T_0)$ and $H_{c2}(T_0)$ are
then changed until, for all fields, the high frequency value for $r_1$
is equal to $b=H/H_{c2}(T_0)$.  Thus, {\it having assumed only the
high frequency limit of} $r_{1}$, $r_{1}(\nu)$ and $r_{2}(\nu)$ at
various fields and temperatures are obtained.  Since
$\frac{\lambda(B,T_{0})}{\delta_{nf}(B,T_{0})}$ has already been
selfconsistently determined at each temperature $T_{0}$, we can
isolate the {\it pure vortex motion complex resistivity}
$\mathrm{Re}[\rho_{v}(\nu)]$ and $\mathrm{Im}[\rho_{v}(\nu)]$ at
various fields and temperatures.  Those curves can then be fitted to
Eq.(\ref{CCformulae}) to get the vortex parameters.

\begin{center}
{\bf Acknowledgments}
\end{center}
This work has been partially supported by INFM under the national
projects PRA-H.O.P and PRA-U.M.B.R.A., and by MIUR under a FIRB
project ``Strutture Semiconduttore/Superconduttore per l'elettronica
integrata''.  We thank V. Ferrando and C. Ferdeghini for supplying the
MgB$_{2}$ sample, C. Camerlingo for YBCO sample Y4 and M. Boffa and
A.M. Cucolo for the remaining YBCO and SmBCO samples.  We
acknowledge useful discussions with S. Anlage, M. Ausloos, M.W.
Coffey, R. Fastampa, M. Giura, J. Halbritter, A. Maeda, R. Marcon, D.
Neri, D. Oates, R. W\"ordenweber.
%
%
%
%

\newcommand{\APL}{{\it Appl. Phys. Lett.} }
\newcommand{\CM}{COND-MAT/}
\newcommand{\EL}{{\it Europhys. Lett.} }
\newcommand{\IJMPB}{{\it Int. J. Mod. Phys. B} }
\newcommand{\JAP}{{\it J. Appl. Phys.} }
\newcommand{\JS}{{\it J. Supercond.} }
\newcommand{\MST}{{\it Meas. Sci. Technol.} }
\newcommand{\PhC}{{\it Physica C} }
\newcommand{\PL}{{\it Phys. Lett.} }
\newcommand{\PR}{{\it Phys. Rev.} }
\newcommand{\PRB}{{\it Phys. Rev. B} }
\newcommand{\PRL}{{\it Phys. Rev. Lett.} }
\newcommand{\RMP}{{\it Rev. Mod. Phys.} }
\newcommand{\RPP}{{\it Rep. Prog. Phys.} }
\newcommand{\SSC}{{\it Solid State Commun.} }
\newcommand{\SUST}{{\it Supercond. Sci. Technol.} }
\newcommand{\JLTP}{{\it J. Low Temp. Phys.} }
%
%
%
\newcommand{\CQG}{{\it Class. Quantum Grav.} }
\newcommand{\CTM}{{\it Combust. Theory Modelling\/} }
\newcommand{\DSE}{{\it Distrib. Syst. Engng\/} }
\newcommand{\EJP}{{\it Eur. J. Phys.} }
\newcommand{\HPP}{{\it High Perform. Polym.} }  
\newcommand{\IP}{{\it Inverse Problems\/} }
\newcommand{\JHM}{{\it J. Hard Mater.} }  
\newcommand{\JO}{{\it J. Opt.} }
\newcommand{\JOA}{{\it J. Opt. A: Pure Appl. Opt.} }
\newcommand{\JOB}{{\it J. Opt. B: Quantum Semiclass. Opt.} }
\newcommand{\JPA}{{\it J. Phys. A: Math. Gen.} }
\newcommand{\JPB}{{\it J. Phys. B: At. Mol. Phys.} }      
\newcommand{\jpb}{{\it J. Phys. B: At. Mol. Opt. Phys.} } 
\newcommand{\JPC}{{\it J. Phys. C: Solid State Phys.} }   
\newcommand{\JPCM}{{\it J. Phys.: Condens. Matter\/} }    
\newcommand{\JPD}{{\it J. Phys. D: Appl. Phys.} }
\newcommand{\JPE}{{\it J. Phys. E: Sci. Instrum.} }
\newcommand{\JPF}{{\it J. Phys. F: Met. Phys.} }
\newcommand{\JPG}{{\it J. Phys. G: Nucl. Phys.} }         
\newcommand{\jpg}{{\it J. Phys. G: Nucl. Part. Phys.} }   
\newcommand{\MSMSE}{{\it Modelling Simulation Mater. Sci. Eng.} }
\newcommand{\NET}{{\it Network: Comput. Neural Syst.} }
\newcommand{\NJP}{{\it New J. Phys.} }
\newcommand{\NL}{{\it Nonlinearity\/} }
\newcommand{\NT}{{\it Nanotechnology} }
\newcommand{\PAO}{{\it Pure Appl. Optics\/} }
\newcommand{\PM}{{\it Physiol. Meas.} }  
\newcommand{\PMB}{{\it Phys. Med. Biol.} }
\newcommand{\PPCF}{{\it Plasma Phys. Control. Fusion\/} } 
\newcommand{\PSST}{{\it Plasma Sources Sci. Technol.} }
\newcommand{\PUS}{{\it Public Understand. Sci.} }
\newcommand{\QO}{{\it Quantum Opt.} }
\newcommand{\QSO}{{\em Quantum Semiclass. Opt.} }
\newcommand{\SLC}{{\it Sov. Lightwave Commun.} }  
\newcommand{\SST}{{\it Semicond. Sci. Technol.} }
\newcommand{\WRM}{{\it Waves Random Media\/} }
\newcommand{\JMM}{{\it J. Micromech. Microeng.\/} }
%
%
\newcommand{\AC}{{\it Acta Crystallogr.} }
\newcommand{\AM}{{\it Acta Metall.} }
\newcommand{\AP}{{\it Ann. Phys., Lpz.} }
\newcommand{\APNY}{{\it Ann. Phys., NY\/} }
\newcommand{\APP}{{\it Ann. Phys., Paris\/} }
\newcommand{\CJP}{{\it Can. J. Phys.} }
\newcommand{\JCP}{{\it J. Chem. Phys.} }
\newcommand{\JJAP}{{\it Japan. J. Appl. Phys.} }
\newcommand{\JP}{{\it J. Physique\/} }
\newcommand{\JPhCh}{{\it J. Phys. Chem.} }
\newcommand{\JPCS}{{\it J. Phys. Chem. Solids} }
\newcommand{\JMMM}{{\it J. Magn. Magn. Mater.} }
\newcommand{\JMP}{{\it J. Math. Phys.} }
\newcommand{\JOSA}{{\it J. Opt. Soc. Am.} }
\newcommand{\JPSJ}{{\it J. Phys. Soc. Japan\/} }
\newcommand{\JQSRT}{{\it J. Quant. Spectrosc. Radiat. Transfer\/} }
\newcommand{\NC}{{\it Nuovo Cimento\/} }
\newcommand{\NIM}{{\it Nucl. Instrum. Methods\/} }
\newcommand{\NP}{{\it Nucl. Phys.} }
\newcommand{\PRS}{{\it Proc. R. Soc.} }
\newcommand{\PS}{{\it Phys. Scr.} }
\newcommand{\PSS}{{\it Phys. Status Solidi\/} }
\newcommand{\PTRS}{{\it Phil. Trans. R. Soc.} }
\newcommand{\RSI}{{\it Rev. Sci. Instrum.} }
\newcommand{\ZP}{{\it Z. Phys.} }
%
%

%
\end{document}